# Enhanced IR Light Absorption in Group IV-SiGeSn Core-Shell Nanowires

Anis Attiaoui,[1] Stephan Wirth,[2] André-Pierre Blanchard-Dionne,[1] Michel Meunier,[1] J. M. Hartmann,[3] Dan Buca,[2] and Oussama Moutanabbir[1, *]

[1]*Department of Engineering Physics, École Polytechnique de Montréal, Montréal, C.P. 6079, Succ. Centre-Ville, Montréal, Québec, H3C 3A7 Canada*

[2]*Peter Grünberg Institute 9 and JARA-FIT, Forschungszentrum Juelich, 52425 Juelich, Germany*

[3]*Université Grenoble Alpes& CEA, LETI, Minatec Campus, 17 rue des Martyrs, 38054 Grenoble, France*

## Abstract,

Sn-containing Si and Ge alloys belong to an emerging family of semiconductors with the potential to impact group IV semiconductor devices. Indeed, the ability to independently engineer both lattice parameter and band gap holds the premise to develop enhanced or novel photonic, optoelectronic, and electronic devices. With this perspective, we present detailed investigations of the influence of $Ge_{1-y-x}Si_xSn_y$ layers on the optical properties of Si- and Ge-based heterostructures and nanowires. We found that adding a thin $Ge_{1-x-y}Si_xSn_y$ capping layer on Si or Ge greatly enhances light absorption especially in the near IR range leading to an increase in short-circuit current density. For the $Ge_{1-y-x}Si_xSn_y$ structure at thicknesses below 30 nm, a 14-fold increase in the short-circuit current is predicted with respect to bare Si. This enhancement decreases by reducing the capping layer thickness. Conversely, decreasing the shell thickness was found to improve the short-circuit current in $Si/Ge_{1-y-x}Si_xSn_y$ and $Ge/Ge_{1-y-x}Si_xSn_y$ core/shell nanowires. The optical absorption becomes very important when increasing the Sn content. Moreover, by exploiting optical antenna effect, these nanowires show an extreme light absorption reaching an enhancement factor, with respect to Si or Ge nanowires, on the order of $\sim 10^4$ in $Si/Ge_{0.84}Si_{0.04}Sn_{0.12}$ and $\sim 12$ in $Ge/Ge_{0.84}Si_{0.04}Sn_{0.12}$ core/shell nanowires. Furthermore, we analyzed the optical response of the addition of a dielectric capping layer consisting of $Si_3N_4$ to the $Si/Ge_{1-y-x}Si_xSn_y$ core-shell nanowire and found about 50% increase in short-circuit current density for a dielectric layer thickness of 45 nm and a core radius and shell thickness superior to 40 nm. The core/shell optical antenna benefits from a multiplication of enhancements contributed by leaky mode resonances in the semiconductor part and antireflection effects in the dielectric part.

PACS: 81.07.-Gf, 78.67.-Uh, 78.20.-Ci, 07.57.-Kp, 78.66.-w,

# I. Introduction

Engineering light absorption and scattering in group IV semiconductors has been a powerful paradigm to implement innovative and high performance devices for solar cells [1], photodiodes, photodetectors, and plasmonics [2,3], to name a few. Of particular interest, developing cost-effective and high efficiency photovoltaic cells relies crucially on the availability of affordable materials that can efficiently absorb light and convert it into charge carriers. Within this broad context, silicon (Si) has been the material of choice due to its cost-effectiveness and its wide use in semiconductor technologies. However, despite its attractive electronic and material characteristics, Si has rather poor optical properties, relative to its mainstream semiconductor counterparts, due to its indirect band gap and limited light absorption especially in near-infrared region (NIR) and beyond thus limiting the efficiency of Si-based solar cells. Consequently, electricity generated by Si solar cells, which currently makes up over 90% of the photovoltaic market, is still 2-3 times more expensive than that generated from conventional fossil fuels which hinders their competitiveness and wide-scale adoption [4]. In order to enhance the performance of Si-based cells, a variety of strategies for light trapping through surface texturing and coating have been recently proposed [5,6]. Among these strategies, nanoscale structures such as nanowires (NWs) are emerging as effective building blocks to enhance Si optical properties to enable a broader range of optoelectronic devices [7–9], such as photovoltaic cells [10–13], photodetectors [14–16], meta-materials [17–19] and thermal emitters [20]. In this work, our calculations demonstrate that Sn-containing group IV ($Ge_{1-x-y}Si_xSn_y$) core-shell nanowire (CSNW) structures should be effective in enhancing NIR light absorption in Si-based structures. We also demonstrate that this additional Sn-containing shell enhances the optical properties of germanium nanowires (GeNWs) as well.

Ge$_{1-x-y}$Si$_x$Sn$_y$ is an emerging semiconductor system providing two degrees of freedom for band structure engineering, namely, alloying and strain. This ability to independently manipulate strain and lattice parameter is central to engineer novel group IV low-dimensional systems and heterostructures in a similar fashion to the more mature III-V materials. Moreover, unlike the indirect band gap Si and Ge, a direct band gap semiconductor can be achieved using Sn-containing alloys thus providing a promising path towards Si-compatible efficient devices for emission and detection of light [21,22]. The nature of band gap and its energy depend on the composition of the alloys [23–26]. In the following, we begin by studying the structural properties and crystalline quality of Ge$_{1-x-y}$Si$_x$Sn$_y$ layers at various compositions. Next, the optical properties of these layers are investigated using spectroscopic ellipsometry. The obtained optical properties of the analyzed layers were introduced to a Mie-scattering formalism using cylindrical CSNWs to theoretically evaluate the effects of the dimensions of a NW on the scattering and absorption efficiencies. We start by demonstrating the optical antenna effects in individual SiNWs and GeNWs, which form the basic building blocks of our proposed CSNWs structures. We then continue with the optimization of their absorption efficiencies by engineering the best possible match between the absorption spectrum of the wires and visible range (0.4-0.75 μm) as well as the NIR range. Finally, as a figure of merit of absorption efficiency, we analyze the photocurrent of different group IV Ge$_{1-x-y}$Si$_x$Sn$_y$ based films and NW-based structures: Si/Ge$_{1-x-y}$Si$_x$Sn$_y$ CSNWs, Ge/Ge$_{1-x-y}$Si$_x$Sn$_y$ CSNWs, and CSNWs wrapped in non-absorbing, anti-reflection coating (ARC) layers. Using these structures, we obtained extremely-large absorption and scattering enhancements as compared to SiNWs, GeNWs, and thin films.

# II. Structural and Optical Properties

Detailed investigations of structural and optical properties of $Ge_{1-x-y}Si_xSn_y$ semiconductors are still in their infancy despite their scientific and technological importance. Herein, in order to extract the optical properties needed for a more accurate theoretical treatment of light-NW interaction, we employed spectroscopic ellipsometry to characterize $Ge_{1-x-y}Si_xSn_y$ thin layers. Figs. 1 exhibits high resolution Scanning Transmission Electron Microscopy (STEM) images near the $Ge_{1-x-y}Si_xSn_y$/Ge interface for two different Sn and Si compositions [(x, y) = (12%, 4%) and (4%, 12%), respectively]. Details of the growth are provided in Methods section. Also shown in Fig. 1 are the corresponding EDX maps of the three elements Si, Ge, and Sn in two representative ternary layers. The data displayed in Fig. 1 clearly indicate the absence of dislocations or extended defects in the ternary layers or at the epitaxially sharp interface with the Ge VS underneath. X-ray reciprocal space mapping (not shown) confirmed that the grown layers are strained with an out-of-plane tetragonal distortion, typical to pseudomorphic growth, in agreement with electron diffraction patterns (insets in Figs. 1(a) & (b)). The latter also demonstrate the absence of Sn precipitates despite that the Sn content exceeds by several factors its equilibrium solubility (<1.0 at. %).

Next, the optical properties of these single crystalline, metastable layers were investigated by room temperature Variable-Angle Spectroscopic Ellipsometer (VASE). Using the visible-UV instrument, the dielectric function of different samples was determined from 1 to 5 eV with 0.01 eV steps based on measurements at four different angles of incidence. The ellipsometric data processing is elaborated in the supplementary information (SI, section S1). After extracting the optical constants, we incorporate the complex refractive index into a Lorentz-Mie formalism in order to quantify the absorption, scattering, and extinction efficiencies of CSNW, as it will be showed in the following section. The $N(E)$ spectra resulting from the modeling of $\Psi$ and $\Delta$, are displayed in Figure 2. In this figure, we present the obtained spectra for the binary alloy semiconductor Ge$_{1-y}$Sn$_y$ (top) and the ternary alloy Ge$_{1-x-y}$Si$_x$Sn$_y$ (bottom), where the composition of Sn has been measured by EDX and Rutherford backscattering spectrometry [27].

## III. Theoretical Calculations

The calculation of the optical absorption of the core-shell nanowire structure are performed within the framework of the Lorentz-Mie scattering formalism [28] adapted to a cylindrical geometry. The CSNW is modeled as an infinitely long coaxial cylinder (the diameter is at least 10 times smaller than the length), which has been extensively used to analyze light absorption in various NWs [29–32]. The CSNW has a total radius of $t + R_c$, where $t$ and $R_c$ are the shell thickness and the core radius, respectively. The nanowire, placed in vacuum, is illuminated by a plane wave, the wavevector of which, $\boldsymbol{k}_{in}$, is perpendicular to the NW axis, as depicted in Fig. S1 in SI. Furthermore, we have fixed the wavelength range to 300-1100 nm covering the visible and NIR range of the solar spectrum (AM1.5G) [33]. The complex refractive index of Ge$_{1-y}$Sn$_y$ and Ge$_{1-x-y}$Si$_x$Sn$_y$ ternary alloys were extracted from spectroscopic ellipsometry measurements outlined above. Here, we will focus on the Sn-rich Ge$_{0.88}$Sn$_{0.12}$ and Ge$_{0.84}$Si$_{0.04}$Sn$_{0.12}$ alloys because their composition is very close to those predicted for direct band gap semiconductors with a direct gap energy of 0.49 eV [34] and 0.46 eV [23], respectively. The NW far-field optical response is characterized by two dimensionless quantities: the absorption efficiency $Q_{abs}$ and the scattering efficiency $Q_{sca}$ as shown below:

$$Q_{abs(scat)} = \frac{Q_{abs(sca)}^{TM} + Q_{abs(sca)}^{TE}}{2} \qquad (1)$$

where $Q_{abs(sca)}^{TE}$ and $Q_{abs(sca)}^{TM}$ are the absorption (scattering) efficiencies for transverse electrical TE (electric field is perpendicular to the NW axis) and transverse magnetic TM (electric field is parallel to the NW axis) polarisation, respectively.

These quantities correspond respectively to the ratio of the absorption ($C_{abs}$) and scattering ($C_{sca}$) cross section to the geometrical cross section $C_{geo}$ of the NW [28]. The cross section is defined as a fictive area around the illuminated object. As soon as a light beam hits this area, the interaction occurs. Let us consider an ideal case where the absorption cross section $C_{abs}$ of a hypothetical, perfectly absorbing black body is always equal to the geometrical area $A$ of the object, the absorption efficiency is therefore always equal to 1. Now, for a system that is an imperfect absorber, the absorption efficiency $Q_{abs}$ is between 0 and 1 within the limits of geometrical optics. However, when a structure is smaller than the illuminating wavelength $\lambda$, such as the case for the studied NWs (where the diameters are typically below 200 nm), the absorption efficiencies $Q_{abs}$ can exceed unity [28]. This can be interpreted by higher absorption cross section $C_{abs}$ as compared to the geometrical cross section $C_{geo}$. In other words, the NW can collect light from an area much bigger than its geometrical area $A$ [12,29,35,36].

In the following, we briefly present the key equations under the Lorenz-Mie framework allowing the quantification of light absorption by CSNW structures. Under unpolarized illumination, the NW far-field optical response is the average of the absorption efficiencies between TE and TM modes and it is given by Eq. 1.

In addition, the absorption efficiency is deduced from the difference of the extinction and scattering efficiencies. Thus, we have:

$$Q_{abs}^{TM} = Q_{ext}^{TM} - Q_{sca}^{TM}; \qquad Q_{abs}^{TE} = Q_{ext}^{TE} - Q_{sca}^{TE}; \qquad (2)$$

These efficiencies are explicitly given by:

$$Q_{ext}^{TE} = \frac{2}{rk_0} Re\left\{\sum_{n=-\infty}^{\infty} a_n\right\}; \qquad Q_{ext}^{TM} = \frac{2}{rk_0} Re\left\{\sum_{n=-\infty}^{\infty} b_n\right\};$$

$$Q_{sca}^{TE} = \frac{2}{rk_0} Re\left\{\sum_{n=-\infty}^{\infty} |a_n|^2\right\}; \qquad Q_{sca}^{TM} = \frac{2}{rk_0} Re\left\{\sum_{n=-\infty}^{\infty} |b_n|^2\right\}; \qquad (3)$$

where $r$ is the radial dimension of a NW, which is equal to $R_c + t$. Note that the NW core and the shell are composed of different semiconductors that can both contribute to light absorption. Finally, $a_n$ and $b_n$ can be readily obtained by solving Maxwell's equations with the appropriate boundary conditions [28,37] at the core/shell interface and shell/air interfaces. We present in SI, Section S2, a new detailed treatment for the TE and TM modes for a random incidence angle. In addition, to benchmark the Lorentz-Mie scattering code, we validated our calculations by studying the scattering efficiency of SiNWs and comparing our results with what have been experimentally reported in literature. The full benchmark study is presented in SI, Section S3.

# IV. Results and Discussion

We investigated different CSNWs, including Si/Ge$_{1-y}$Sn$_y$, Si/Ge$_{1-x-y}$Si$_x$Sn$_y$, Ge/Ge$_{1-y}$Sn$_y$ and Ge/Ge$_{1-x-y}$Si$_x$Sn$_y$. First, The Lorentz-Mie calculation is used in order to generate 2D absorption and scattering efficiencies maps as a function of the incident wavelength and the NW core radius for the Si/GeSn and Ge/GeSn structures. We start by examining the absorption and scattering efficiency maps of CSNWs with a shell made of the binary alloy Ge$_{1-y}$Sn$_y$ and a core made of Si or Ge at a shell thickness $t = R_c/4$ and $t = R_c$ (Figs. 3 & 4). Fig. 3 exhibits the absorption (top) and scattering (bottom) efficiencies of the Si/Ge$_{0.88}$Sn$_{0.12}$ CSNW, whereas Fig. 4 shows the corresponding results for the Ge/Ge$_{0.88}$Sn$_{0.12}$ CSNW. Moreover, for comparison, the figures also display the absorption and scattering efficiencies for SiNW and GeNW thus highlighting the effect of the Sn-containing shell on the absorption and scattering efficiencies. For wavelength larger than ~550 nm, $Q_{sca}$ shows distinct features in terms of bright whiskers for both TE and TM polarization (not shown in Figs. 3&4, but can easily be inferred from the unpolarised 2D map because $Q_{sca}^{Unp} = (Q_{sca}^{TE} + Q_{sca}^{TM})/2$). The slope of the whiskers is decreasing with increasing core radius $R_c$ and decreasing wavelengths. A possible way to explain these whiskers of the scattering efficiency is by considering the theoretical expression of $Q_{sca}$ given by the Lorenz-Mie formalism for a CSNW (formulas of the scattering coefficients in SI, section 2). In fact, since the complex refractive index of Ge$_{0.88}$Sn$_{0.12}$ shows a weak dependence on wavelength for λ>550 nm (see Fig. 2), the scattering coefficients are proportional to the size parameter $x_{j\ (j\in\{shell,core\})} \propto R_c/\lambda$. Thus, $Q_{sca}$ is nearly constants along straight lines in the $(\lambda, R_c)$ –plane. The same behavior is observed for scattering efficiency $Q_{sca}$ for both CSNW systems with a Ge$_{0.88}$Sn$_{0.12}$ shell but the whiskers are observed in different regions.

Next, we study the effect of the shell thickness on the absorption and scattering efficiencies. By examining the data obtained for the Si/Ge$_{0.88}$Sn$_{0.12}$ CSNW (Fig. 3), two important observations emerge. First, the branched resonance distribution shows an evolution as a function of the core and shell dimensions. In particular, there is a redshift in the whiskers distribution with increasing GeSn shell thickness $t$. Indeed, at fixed core radius (e.g., $R_c = 60\ nm$) and an increasing shell thickness, Figure S3(a, b) displays qualitatively this behaviour where the effect of the geometrical parameters on the leaky resonant modes is shown. In fact, by focusing on the TM$_{41}$, we can deduce that increasing the shell thickness will induce a redshift in the leaky mode resonance. Besides, the higher leaky modes are located at smaller wavelengths, whereas the fundamental mode $TM_{01}$ is around 1000 nm. Three resonant Mie absorption peaks were observed at 720.6 nm, 932 nm, 733 nm, and 1073 nm for a shell thickness equal to $0.5 \times R_c$ (green curve in Fig. S3(a)). Incident electromagnetic waves having specific wavelength, λ, are trapped along the periphery of core/shell NWs similar to whispering gallery modes in micrometer-scale resonators. The resonant field intensity is built up inside the NW and then the confined mode leaks due to the small size of NWs compared to the wavelength of the light ($\lambda > 300\ nm$ and $d < 200\ nm$). The leakage effect is observable in the electric field profile shown in Fig. S3 (b) for the fundamental leaky mode $TM_{01}$ and a core radius of 8 nm and a shell thickness equal to $0.25 \times R_c$ .

In order to quantitatively analyze the leaky modes evolution *vs.* the core radius, we numerically solved the Helmholtz Eigen-equation by using a full-vector finite difference (FVFD) approach coupled with a perfectly matched layer (PML) boundary condition terminated with zero boundary condition [38] to find the complex effective refractive index $n_{eff}$ of the CSNW. Each complex solution is the eigenvalue of a specific leaky mode. Leaky modes can be characterized by an azimuthal mode number, *m*, which indicates an effective number of wavelengths around the wire circumference and a radial order number, *l*, describing the number of radial field maxima within the cylinder (for instance TM*ml*) [35]. The leaky mode resonance, TE*ml* and TM*ml*, have previously been shown to correspond to peaks in the scattering and absorption spectra of NWs [35,39]. The real part of $n_{eff}$ is indicative of the resonant wavelength and propagation constant and the imaginary part is indicative of the radiative loss of the mode, which for a lossy medium is also a measure of the absorptive loss. Thus, we can estimate the resonant scattering wavelength from $m\lambda/n_{eff} = \pi d$ where $m = 1, 2, ...$, $\lambda$ is the free space wavelength of incident light, $n_{eff}$ is the effective refractive index, and $d$ is the diameter of the nanowire. For instance, the nanowire with 120 nm diameter ($R_c = 60\ nm$) exhibits an absorption peak at 1073 nm as shown in Fig. S3(b). By using the above formula, we obtain $n_{eff} = 2.862$ (for m = 1), which is close to the refractive index of Si core (3.55) at this wavelength. Using the FVFD approach, we estimate the effective refractive index to be 3.13656.  Such result reveals two important features: the optical antenna effect is maximized at each resonance and the scattering and absorption spectra are highly structured with multiple discrete peaks. This optical antenna effect is enhanced by decreasing the NW core radius. As shown in Figs. 3 and 4 (bottom), when the core radius is smaller or the wavelength of the incident light is longer, the scattering efficiency increases. The data presented in Figs. 3 and 4 clearly demonstrate the ability to tune the absorption and scattering of light using CSNWs by controlling either the NW radius or the light wavelength in the visible and especially the NIR range where the absorption and scattering efficiencies are significantly high.

The second important element affecting the absorption efficiency is the presence of localized resonant modes. Similar observations were reported in different systems: GaN nanowire cavity [40], hydrogenated amorphous silicon a-Si:H core with a dielectric shell [41] and polycarbonate (PC) - polyvinylidene difluoride (PVDF) CSNW [42]. Interestingly, it is possible to engineer the resonant property inside a CSNW by tuning the core radius and/or the shell thickness in a way that light absorption can be enhanced at resonance regions, the so called the leaky-mode resonance (LMR) enhancement. This effect will be explored in order to optimize the morphology of CSNWs to achieve an efficient light absorption. For this, in the following we define the optimal core radius, shell thickness, and the shell type (be it Ge$_{1-y}$Sn$_y$ binary alloy or Ge$_{1-x-y}$Si$_x$Sn$_y$ ternary alloy). The absorption behavior can be understood by means of Fano-resonance effect [43] that is an interference effect arising from the incident light and the localized reemitted LMR light due to the subwavelength size of NWs. Moreover, in order to quantify the absorption of different CSNW structures across the solar spectrum, we calculate the ultimate efficiency $\eta_{eff}$ [44] defined assuming that each absorbed photon with energy greater than the band gap produces a single electron-hole pair with energy $hc/\lambda_g$, where $\lambda_g$ is the wavelength corresponding to the minimum band gap between the core and the shell. Note that unstrained Si and Ge have an indirect band gap energy of 1.12 eV and 0.66 eV, respectively, whereas the unstrained Ge$_{0.88}$Sn$_{0.12}$ is a direct gap semiconductor having a gap of 0.49 eV [23]. $\eta_{eff}$ is given by:

$$\eta_{eff} = \frac{\int_{0.3\,\mu m}^{\lambda_g} F_s(\lambda) Q_{abs}(\lambda, r_c, t) \frac{\lambda}{\lambda_g} d\lambda}{\int_{0.3\,\mu m}^{4\,\mu m} F_s(\lambda) d\lambda};$$

(4)

where $\lambda$ is the wavelength, $F_s(\lambda)$ is the spectral irradiance (power density in W.m$^{-2}$.nm$^{-1}$) of the ASTM AM1.5G direct normal and circumsolar spectrum [33], $Q_{abs}(\lambda, R_c, t)$ is the spectral absorption efficiency evaluated with the Mie-Lorentz scattering formalism and $\lambda_g$ is the wavelength corresponding to the smallest band gap between the core and the shell. Special care is needed when evaluating the integrals in Eq. 4, due to the non-uniformity of the spectral range in the ASTM data. To solve this issue, we simply used a Piecewise cubic Hermite interpolating polynomial to interpolate the optical properties between 300 and 1100 nm, to match the ASTM spectra wavelength steps. For instance, if we consider the Si/GeSn CSNW, then $\lambda_g = 2.583$ μm. The ultimate efficiency can be linked to the maximum short-circuit current density, $J_{sc}$, by assuming an ideal carrier collection efficiency where every photogenerated carrier reaches the electrodes and contributes to photocurrent. Under this condition, we write:

$$J_{sc} = \frac{q}{hc} \int_{0.3\ \mu m}^{\lambda_g} F_s(\lambda) Q_{abs}(\lambda, R_c, t) \lambda d\lambda = 0.0726 \times \lambda_g \eta_{eff}\ [mA/cm^2]$$

(5)

where $q$ is the elementary charge. However, because $Q_{abs}$ can reach values greater than unity, as explained above, the integrated solar absorption cannot truly be considered as the real ultimate photocurrent intensity or the short-circuit current density. Nevertheless, $J_{sc}$ is a figure of merit proportional to the actual photocurrent intensity, very useful to compare the absorption efficiency of the studied quantum structures [13].

The photocurrent enhancement is defined as the ratio of the short-circuit current of a two-layer stack to the short-current of a Si or Ge film. For instance, the enhancement of the photocurrent of Si/Ge$_{0.88}$Sn$_{0.12}$ (blue solid line in Fig. 5) is evaluated by calculating the ratio $\eta_{J_{sc}}$, given by:

$$\eta_{J_{sc}} = \frac{J_{sc}^{Si/GeSn}}{J_{sc}^{SiNW}}$$

(6)

where $J_{sc}^{Si/GeSn}$ and $J_{sc}^{SiNW}$ are the photocurrents generated in Si/Ge$_{0.88}$Sn$_{0.12}$ stack and Si thin film, respectively. $\eta_{J_{sc}}$ displays a few peaks attributed to Fabry-Pérot resonance which is much weaker than LMR in NWs. Before investigating the enhancement of the absorption in CSNWs, we present in Fig. 5 the calculated short-circuit current density enhancement of a bilayer thin film structure of equal thickness $d_1 = d_2$ composed of different group IV semiconductor alloys: Si/ Ge$_{0.88}$Sn$_{0.12}$, Ge/ Ge$_{0.88}$Sn$_{0.12}$, Si/ Ge$_{1-x-y}$Si$_x$Sn$_y$ and Ge/ Ge$_{1-x-y}$Si$_x$Sn$_y$ with (x, y) = (12%, 4%) and (4%, 12%). The current density was evaluated using Eq. 5 and the absorption efficiency for the single and double layer was calculated using the Transfer Matrix Method (TMM) described in full detail in Reference [45] under a TM polarization and normal incidence. Fig. 5 demonstrates a significant enhancement in the photocurrent when Si and Ge are capped by an Sn-containing binary or ternary layer. This enhancement is much more pronounced for Si-based layer where a ~9 nm-thick Ge$_{0.88}$Sn$_{0.12}$ or Ge$_{0.84}$Si$_{0.04}$Sn$_{0.12}$ capping layer yields a ~14-fold increase in photocurrent as compared to bare Si. Moreover, we can conclude by comparing the short-current enhancement of Si, Ge, and Si/Ge$_{0.99}$Sn$_{0.12}$ NWs that these large enhancement factors are a result of the optical properties of the alloy, while its narrower bandgap seems to have a limited effect. For a Ge substrate, adding Sn-rich binary or ternary layers (Ge$_{0.88}$Sn$_{0.12}$ or Ge$_{0.84}$Si$_{0.04}$Sn$_{0.12}$) only leads to ~25% an increase in the photocurrent at an optimal thickness of 11 nm. It is also important to note that the richer in Sn the layers are, the higher the short-circuit current enhancement is. We will investigate later on if this behavior still holds true for CSNW structures and compare the photocurrent enhancement of CSNW to that of thin films. It is worth mentioning that in a planar structure, increased reflection or backscattering from the material's front surface decreases light absorption. Thus, changing the geometry from thin films to quasi 1D subwavelength NW would increase scattering indicative of an enhanced optical antenna effect, which increases both scattering and absorption in the NW.

To investigate the absorption enhancement in CSNWs, we evaluate the evolution of the photocurrent enhancement as a function of the NW dimensions in the four structures proposed above (Fig. 6). Fig. 6(a) displays the variation of the ratio $\eta_{J_{sc}}$ as a function of the shell thickness ($t$) and the core radius ($R_c$). We recognize the whiskered features, similar to the absorption efficiency, presented in Figs. 3 and 4, attributed to LMR modes. Interestingly, the 2D photocurrent enhancement maps of Si/Ge$_{0.88}$Sn$_{0.12}$ and Si/Ge$_{0.84}$Si$_{0.04}$Sn$_{0.12}$ NW structures show narrow regions ($R_c < 56 \, nm$ and $t < 40 \, nm$) where an extreme enhancement of $J_{sc}$ is achieved reaching an increase of 13- to 22-fold for Si/Ge$_{0.88}$Sn$_{0.12}$ and 22- to 47-fold for Si/Ge$_{0.84}$Si$_{0.04}$Sn$_{0.12}$ relative to SiNW. Besides, from the $\eta_{J_{sc}}$ maps of Ge/Ge$_{0.88}$Sn$_{0.12}$ and Ge/Ge$_{0.84}$Si$_{0.04}$Sn$_{0.12}$, when the core radius is larger than the shell thickness ($R_c > t$), we observe a maximum photocurrent enhancement of 7 fold compared to GeNW system. The enhancement is governed by the LMR modes, where the largest increase occurs for a shell thickness below 8 nm and a core radius between 30 and 45 nm. Table 1 summarizes the key results of the short-circuit current enhancement for the above CSNW structures.

Additionally, in order to quantitatively compare the photocurrent enhancement between CSNW and thin film systems, we present in Fig. 6(b) the ratio $\eta_{J_{sc}}$ for the Si/Ge$_{0.88}$Sn$_{0.12}$ at two different shell thicknesses ($t = R_c/4 \, or \, R_c$) for the CSNW (red curves) and two top layer thicknesses ($d_2 = d_1/4 \, or \, d_1$) for the thin films (blue curves). The short-current enhancement $\eta_{J_{sc}}$ is evaluated by calculating the ratio of $J_{sc}$ of the CSNW structure with a core radius $R_c$ and a shell thickness $t$ by $J_{sc}$ of a NW with a radius of $R_c + t$. We note that $\eta_{J_{sc}}$ of the CSNW at $t = R_c$ is almost comparable to its counterparts for the thin film. Nevertheless, when we decrease the shell thickness from $t = R_c$ to $t = R_c/4$ for the CSNW and from $d_2 = d_1$ to $d_2 = d_1/4$ for the thin film, we notice that the maximum of $\eta_{J_{sc}}$ increases from 14 to 19 in the CSNW and decreases from 13 to 8 in the thin film. This is an interesting finding suggesting that in order to increase light absorption in CSNW, the shell thickness $t$ of the CSNW has to be thinner than the core radius $R_c$. Thus, on the one hand, decreasing the shell thickness from $R_c$ to $R_c/4$ of Si/GeSn CSNW will improve-in average- the short-circuit current by 45%. On the other hand, decreasing the top layer thickness in the Si/GeSn thin-film from $d_1$ to $d_1/4$ will deteriorate-in average- the short-circuit current by 15%.

To better illustrate the effect of shell thickness on the generated photocurrent, data obtained for three different thicknesses ($t = 3, 50,$ and $100$ nm) are selected from the 2D photocurrent maps (dashed lines labeled a, b and c in Fig. 6(a)) and plotted in Fig. 7. The corresponding photocurrents are compared to those generated in SiNW and GeNW. Note that the latter have been extensively investigated in recent years [29,46–48] and our estimated photocurrents for bare SiNW and GeNW are in full agreement with these early reports. More importantly, Fig. 7 clearly shows that the addition of a SiGeSn or GeSn shell enhances significantly the photocurrent. For instance, the photocurrent enhancement has almost increased 10-fold in Si/GeSn and 20-fold in Si/GeSiSn CSNW structures as compared to SiNW (black curves in Fig. 7) at a shell thickness of 3 nm and core radii larger than 60 nm. Also, when the shell thickness is equal to 3 nm (Fig. 7(a)), we obtain the highest value for $J_{sc}$ reaching up to 130 mA/cm$^2$ for the Ge/GeSn structure at a core radius of 40 nm. This is due to the presence of the TM$_{11}$/TE$_{01}$ leaky resonance mode at a core radius above 30 nm for all the studied NWs. We also notice an important improvement in $J_{sc}$ at smaller core radius which can reach 120 mA/cm$^2$ for Ge/GeSn, Ge/GeSiSn and Si/GeSiSn structures for a 3 nm-thick shell. Increasing the shell thickness will contribute to deterioration of the short-current, as it can be seen from Fig. 7(b) and Fig. 7(c) and has been proven from the aforementioned analysis of Fig. 6b. Additionally, the comparison of panels (b), and (c) in Fig. 7 shows that there is relatively little difference between absolute values, suggesting similar behavior for shell thickness larger than 50 nm. This can be deduced from top left panel in Fig. 6a. This extreme enhancement at a very small core radius ($R_c = 7\ nm$) and shell thickness ($t = 3\ nm$) is due to the TM$_{01}$ leaky resonance mode which is proven by the near-field profile in Fig. 7.

The fact that the photocurrent enhancement is practically independent of the core radius above ~100 nm is an interesting result suggesting that the generated photocurrent can be tuned through a simple control of the shell thickness rather than the core radius thus providing more flexibility in the fabrication process. However, to ensure the integration of these structures in photonic devices, it is of paramount importance to also optimize the efficiency of charge collection, which depends on minority carrier lifetime, lattice defects, and contact quality and design.

In addition, to further investigate the optimal CSNW geometry for light absorption, we defined a dimensionless parameter, $\eta_{abs}$, as the ratio of the absorption efficiency of the CSNW to that of a pure NW (SiNW or GeNW) with the same dimension as its CSNW counterparts. For instance, if we consider the structure Si/GeSn, the absorption enhancement factor is defined as

$$\eta_{abs}(r_c, t, \lambda) = \frac{Q_{abs}^{Si/GeSn}}{Q_{abs}^{SiNW}};$$ 

(7)

If $\eta_{abs}$ takes values larger than unity, then we conclude that light is absorbed efficiently by the CSNW structure as compared to SiNW or GeNW. The higher the value of $\eta_{abs}$, the more efficient absorber the CSNW is.

The enhancement of absorption and photocurrent are due to the plurality of spectrally-separated LMR supported by large core diameters. In the enhancement maps (Fig. 8 for Si/Ge$_{0.88}$Sn$_{0.12}$, Fig. S4, SI for Si/Ge$_{0.84}$Si$_{0.04}$Sn$_{0.12}$ and Fig. S5, SI for Ge/Ge$_{0.88}$Sn$_{0.12}$), a strong absorption enhancement that follow very specific directions is observed. In fact, when the incident wavelength matches one of the leaky modes supported by the NW [46,47], optical responses, including light scattering and absorption, were found to be substantially enhanced as compared to SiNW. We observe a gradual increase in the enhancement of light absorption for the Si/GeSn system compared to SiNW where between 800 and 1000 nm, $\eta_{abs}$ ~$10^3$. Next, between 1000 and 1100 nm, $\eta_{abs}$ reaches its maximum at ~$10^4$. For instance, at $r_c = 33\ nm$, $\lambda = 1086\ nm$ and $t = R_c$, $\eta_{abs} \cong 3 \times 10^4$ for Si/GeSn CSNW. Additionally, when increasing the thickness of the shell, leakier resonance modes are excited, which is reflected by the increase in the number of whiskers in the map: at $t = R_c/4$, only five broad whiskers are observed, whereas at $t = R_c$, eight are found for Si/Ge$_{0.88}$Sn$_{0.12}$ at identical core radius. Interestingly, the Ge/Ge$_{0.88}$Sn$_{0.12}$ CSNW's enhancement map (see Fig. S4 in SI) displays a different result. Indeed, unlike Si-based CSNW, the enhancement in Ge-based CSNW is relatively small (~12 *vs.* ~$10^4$). Besides, above 700 nm, the enhancement along the LMR is clear and it reaches its maximum in the NIR region for $860 \leq \lambda \leq 1100\ nm$ and $R_c = 42, 37, 32\ and\ 30\ nm$ for each shell thickness $t$. Specifically, at 980 <$\lambda$< 1100 nm, $R_c = 30\ nm$ and $t = R_c$, we obtain an absorption enhancement factor $\eta_{abs}$ between 9 and 12. This is in accordance with the enhancement map where at the same core radius and shell thickness, $\eta_{J_{sc}}$ is equal to 3.5 corresponding to current density of 95 mA/cm$^2$.

The extreme enhancement of light absorption in Si/GeSn and Si/GeSiSn structures presents an interesting opportunity to achieve optimal wavelength selectivity in the desired NIR region through an optimal choice of the core radius and shell thickness of the CSNW. We therefore need to optimize the physical dimensions ($R_c$, $t$) of the CSNW in order to guarantee both photocurrent and absorption efficiency enhancement. Thus, for Si/Ge$_{0.88}$Sn$_{0.12}$, we propose core radii $R_c =$17, 41, 70, 113, 132 and 172 nm for a shell thickness $t = R_c/4$, at which $Q_{abs}$ is enhanced. In addition, when $R_c = 6.3, 17$ and 38 nm, the current is enhanced as shown in the red solid curve of Fig. 6(b) where the red arrows indicate the corresponding core radii. Thus, a core-radius of 17 nm would guaranty a simultaneous photocurrent and absorption efficiency enhancement. Besides, for $t = R_c$, we find that the optimal core radii are 10, 32, 73, 108 and 149 nm. On the other hand, for Si/ Ge$_{0.84}$Si$_{0.04}$Sn$_{0.12}$ and at $t = R_c$, we find the optimal radii to be 9, 31, 50, 72, 106, 118 and 147 nm.

The investigations described above provide the basis to design high performance NW-based optoelectronic and photonic devices. In the following, we address the influence of an additional layer around the CSNW structure and elucidate the collective properties of an array of CSNWs. A single configuration will be considered: a non-absorbing dielectric layer (Si$_3$N$_4$) around an Si/Ge$_{0.88}$Sn$_{0.12}$ CSNW (Si/Ge$_{0.88}$Sn$_{0.12}$/Si$_3$N$_4$). The optical properties of Si$_3$N$_4$ are taken from Ref. [49]. The choice of silicon nitride (Si$_3$N$_4$) as a coating material emanates from two characteristics: firstly, the surface passivation effect [50,51] and secondly the antireflective properties reducing considerably light reflection. Taking into consideration these aspects, we modeled light absorption and conversion in Si/Ge$_{0.88}$Sn$_{0.12}$/Si$_3$N$_4$ NW. Because optical resonances serve to enhance the light−matter interaction of the NW cavity, we expect the dielectric-shell optical antenna effect to increase not only light scattering but also light absorption in PV devices [46]. Figure S1(c) in SI shows the geometrical configuration and the necessary parameters for the aforementioned system. Fig. 9 displays the calculated $\eta_{J_{sc}}$ ratio at a fixed inner-shell ($t$) of the Ge$_{0.88}$Sn$_{0.12}$ layer equal to the core radius ($t = R_c$) and we varied the dielectric capping layer thickness $D$ from 1 to 200 nm. In the top and left panels in Fig. 9(a), we show the relative change -in percent- of the maximum current enhancement, *i.e.* a positive value means an enhancement or increase in the $J_{sc}$, whereas a negative value entails a decrease in $J_{sc}$. Additionally, the antireflection role of the dielectric shell can be further confirmed from examining the absorption spectra. We can see that the dielectric shell of the Si/Ge$_{0.88}$Sn$_{0.12}$/Si$_3$N$_4$ NW gives rise to a broad absorption peak, clearly shown in Fig. 9(b, c) for two distinct core radii (13.6 and 78.2 nm). So, a core radius $R_c$ equal to 78.2 nm (green arrow in top panel in Fig. 9(a)), an inner shell thickness $t$ equal to the core radius and a thickness of the dielectric capping layer $D$ equal to 43 nm (blue arrow in left panel of Fig. 9(a)) provides 25% increase in the short-current density as well as high absorption efficiency. The enhancement at smaller core radius ($R_c < 20\ nm$; more specifically at $R_c = 13.6\ nm$, shown as an orange arrow in top panel of Fig. 9(a) where ~40% increase in $J_{sc}$ is achieved) and at capping layer thickness ($D = 25\ nm$) follows mainly the LMR. This enhancement is explained by an increase in scattering. In fact, in the case of a subwavelength NW, the increase in scattering is indicative of an enhanced optical antenna effect, which increases both scattering and absorption in the NW. This can be observed in Fig. 10(a). Likewise, tuning the spectral range of the absorption throughout the dielectric layer thickness can be accomplished for a given core radius of 13.6 nm: in fact, when $D = 33$ nm, the structure will absorb in the visible spectrum, but when $D = 3$ nm the NIR spectrum will be active, with 40% current enhancement compared to the Si/Ge$_{0.88}$Sn$_{0.12}$ CSNW structure. Furthermore, when the core radius is increased to 78 nm, the optimal $Si_3Ni_4$ capping layer thickness $D$ is found to be 45 nm from Fig. 9(c) with a corresponding short-current enhancement of 25%.

Next, the effect of the additional dielectric layer can be better understood in Fig. 10(a) displaying TM-like mode absorption efficiency ($Q_{abs}^{TM}$) for three sets of NW structures: Si, $Si/Ge_{0.88}Sn_{0.12}$, and $Si/Ge_{0.88}Sn_{0.12}/Si_3N_4$ at $R_c = 13.6$ and $75\ nm$, and $D = 33$ and 45 nm, respectively for each core radius. Fig. 10(b) shows the spatial distribution of the TM polarization of the Poynting vector of the total near-field (see SI, section S2) inside and outside the investigated NWs at wavelengths corresponding to the peaks labeled 1 to 10 in Fig. 10(a). When coating the $Si/Ge_{0.88}Sn_{0.12}$ CSNW with $Si_3N_4$ layer, an increase of the absorption efficiency is apparent regardless of the core radius. Note that adding a GeSn shell and increasing the core radius will induce a redshift in the LMR spectral position where higher orders modes were observed to appear. Furthermore, the number of excited LMR diminishes due to the presence of the dielectric capping layer. In addition, Fig. 10(b) represent the resonant profiles of the total TM-polarized Poynting vector $|\mathbf{S}|_{TM}^2$ inside the nanostructure and display a leaky-mode resonance behavior (see Fig. 10(b) maps 5,10), which is associated with the selective scattering of light in a specific wavelength depending on structure size. The opportunity to control the spatial distribution of the energy flux density in the three aforementioned structures, by introducing a shell layer, provides wavelength tuneability of the absorption efficiency which is clearer in the absorption maps presented in Fig. 9(b, c). Besides, we demonstrate that a simple dielectric shell can double light absorption and dramatically increase light scattering in SiNWs by enhancing the optical antenna effect of the wires. This effect has been proven using a different dielectric layer ($SiO_2$) [52].

# V. Conclusion

In summary, we have demonstrated LMR-induced field enhancements in $Ge_{1-x-y}Si_xSn_y$ alloy heterostructures and core-shell nanowires. Our theoretical calculations demonstrated that a ~14-fold increase in photocurrent can be achieved in $Si/Ge_{1-y-x}Si_xSn_y$ heterostructures as compared to bare Si. Furthermore, when we decrease the outer layer thickness relative to the core radius in the core-shell nanowires, we obtain an increase in the maximum short-circuit current enhancement factor. Conversely, a thinner Sn-containing top layer leads limits the enhancement of light absorption in thin films. Moreover, the photocurrent increase in nanowire is found to be restricted to narrow regions (core radius $R_c < 56\ nm$ and shell thickness $t < 40\ nm$) where a significant enhancement relative to Si nanowires is achieved reaching 11-22 fold for $Si/Ge_{0.88}Sn_{0.12}$ and 25-47 fold for $Si/Ge_{0.84}Si_{0.04}Sn_{0.12}$ core-shell nanowires. Additionally, we showed an extreme enhancement of light absorption for the Si/GeSn, where the absorption efficiency in the near-infrared region is four order of magnitude higher than that of SiNW. For $Ge/Ge_{1-y-x}Si_xSn_y$ core-shell nanowires the enhancement in light absorption is relatively limited as compared to $Si/Ge_{1-y-x}Si_xSn_y$ core-shell nanowires. The observed enhancement is due to a multiplication of contributions from LMRs in both core and shell semiconductors. These effects can be exploited through the control over the size and composition of the nanowire structure. Moreover, the calculations also suggest that the addition of a $Si_3N_4$ ARC layer on $Si/Ge_{1-y-x}Si_xSn_y$ core-shell nanowires improves the absorption efficiency. In fact, by tuning the core radius and the dielectric layer thickness, its possible to selectively control the spectral range (visible or NIR) were the structure becomes optically active. For instance, with a core radius larger than 75 nm, a dielectric layer thickness of 40 nm, a 30% increase in the generated photocurrent relative to $Si/Ge_{1-y-x}Si_xSn_y$ core-shell nanowires can be achieved. The obtained results indicate that Si-based nanowire structures are more advantageous as compared to thin films for applications in solar cells and photodetectors.

## Acknowledgments

OM acknowledges support from NSERC-Canada (Discovery Grants), Canada Research Chair, Fondation de l'École Polytechnique de Montréal, MRIF Québec, Calcul Québec, and Computer Canada. Computations were made on a supercomputer managed by Calcul Québec and Compute Canada. The operation of this supercomputer is funded by the Canada Foundation for Innovation (CFI), ministère de l'Économie, de la Science et de l'Innovation du Québec (MESI) and the Fonds de recherche du Québec - Nature et technologies (FRQ-NT). AA is grateful for technical help from Mr. Bart Oldeman and Calcul Québec- McGill support team.

**Corresponding author**

[*] Email : (O. Moutanabbir) oussama.moutanabbir@polymtl.ca

# Appendix

**RPCVD growth of Ge$_{1-x-y}$Si$_x$Sn$_y$ samples.** These layers were grown using a metal cold-wall reduced pressure chemical vapor deposition AIXTRON TRICENTR® (RP-CVD) for 200/300 mm wafers.[22] The growth of Ge$_{1-x-y}$Si$_x$Sn$_y$ layers was performed on Si (100) wafers using low-defect density Ge virtual substrates.[54–56] The epitaxial layers were grown using Si$_2$H$_6$, Ge$_2$H$_6$ (10% diluted in H$_2$), and SnCl$_4$ precursors using N$_2$ as carrier gas, which warrant reasonable growth rates at temperatures in the 350-475 °C range. The Ge$_{1-x-y}$Si$_x$Sn$_y$ layers were grown with Si and Sn concentrations in the range of 4-20% and 2-12%, respectively. Prior to Raman investigations, the composition and structural properties of Ge$_{1-x-y}$Si$_x$Sn$_y$/Ge/Si layers were characterized using Rutherford backscattering spectrometry (RBS), x-ray reciprocal space mapping (RSM), and transmission electron microscopy (TEM).

**Optical characterisation.** The optical measurements were performed with a commercial ellipsometer (variable-angle spectroscopic ellipsometer J. A. Woollam)[57] with a rotating polarizer, and an auto-retarder that allows us to measure ellipsometric angle between 0˚ and 360˚, and thus to obtain accurate measurement in the spectral regions of small absorption. The angle of incidence was varied using an automatic goniometer stage between 45° and 75° for the GeSn samples and between 70° and 80° for the Ge$_{1-x-y}$Si$_x$Sn$_y$ samples. A fundamental issue for accurate data extraction is the surface state, namely the surface roughness and oxidation. For the grown Ge$_{1-x}$Sn$_x$ and Ge$_{1-x-y}$Si$_x$Sn$_y$ layers, surfaces roughness and oxidation are the main issue. In order to circumvent the oxidation of the GeSi and GeSiSn layers, a chemical treatment consisting of an HCl (38%): DI-H$_2$O wet etch has been done in order to reduce the thickness of the natural formed oxide layers, rinsed in propanol (5 min), and then blown dry with nitrogen gun.

**Numerical Calculation.** In order to implement the Lorentz-Mie scattering formalism, we developed a MATLAB® code to solve the Maxwell equations in all the aforementioned NW systems (NW and CSNW). Furthermore, we developed a Full-Vector Finite-Difference complex mode solver in order to find the effective refractive index of the CSNW structure based on the work presented in Ref. [36]. Calculations were carried out on a PC equipped with a single Intel Core 4 Quad 2.40GHz processor equipped with 32GB of RAM.

# Table Captions

**Table 1:** Core radius and shell thickness range corresponding to the optimal short-current enhancement (the highest value $\eta_{J_{sc}}$) for the four Core/Shell NW structures.

| Core | Si | | Ge | |
|---|---|---|---|---|
| Shell | $Ge_{0.88}Sn_{0.12}$ | $Ge_{0.88}Si_{0.04}Sn_{0.12}$ | $Ge_{0.88}Sn_{0.12}$ | $Ge_{0.88}Si_{0.04}Sn_{0.12}$ |
| $R_c$ (nm) | 3-31 | 3-45 | 30-45 | 30-42 |
| $t$ (nm) | 1-11 | 1-14 | 1-8 | 1-5 |
| $\eta_{J_{sc}}$ | 13-22 | 22-47 | 6 | 5.5 |

# Figure Captions

**Figure 1:** High Angle Annular Dark Field Scanning Transmission Electron Microscopy (HAADF/STEM) image of $Ge_{0.84}Si_{0.12}Sn_{0.04}$ **(a)** $Ge_{0.84}Si_{0.04}Sn_{0.12}$ **(b)** layers grown on Ge virtual substrates. Note the absence of dislocations or extended defects in the ternary layer or at the interface. The corresponding diffraction patterns measured at the interface are shown as inset figures in (a) and (b) confirming the high crystallinity as well as the absence of Sn precipitates. Low magnification HAADF/STEM images and EDX maps of Si, Ge, and Sn in $Ge_{0.84}Si_{0.12}Sn_{0.04}$ **(c)** and $Ge_{0.84}Si_{0.04}Sn_{0.12}$ **(d)** layers.

**Figure 2:** The complex refractive index constant $N(\lambda)$ spectra of (top) $Ge_{1-x}Sn_x$ binary semiconductor alloy as a function of Sn composition and (bottom) of $Ge_{1-x-y}Si_xSn_y$ ternary alloy as function of Si and Sn composition obtained from spectroscopic ellipsometry measurement. The optical properties of Ge (001) substrate (x=0%) in the $Ge_{1-x}Sn_x$ binary are taken from Palik [58]. The inset shows the multilayer model used to extract the optical properties (details in SI, section S1).

**Figure 3:** Unpolarized absorption efficiency $Q_{abs}$ (top) and unpolarized scattering efficiency $Q_{sca}$ (bottom) of Si/$Ge_{0.88}Sn_{0.12}$ CSNW surrounded by air as a function of the core radius $R_c$ and the incident light wavelength for two different shell thicknesses: $t = R_c/4$ and $t = R_c$. For comparison, SiNW absorption and scattering efficiencies are also shown. The three black-dashed lines indicate the selected radii: on-resonance (at $R_c = 8\ nm$) and off-resonance (at $R_c = 60\ nm$).

**Figure 4:** Unpolarized absorption efficiency $Q_{abs}$ (top) and unpolarized scattering efficiency $Q_{sca}$ (bottom) of Ge/$Ge_{0.88}Sn_{0.12}$ CSNW surrounded by air as a function of the core radius $R_c$ and the incident light wavelength for two different shell thicknesses: $t = R_c/4$ and $t = R_c$. For comparison, GeNW absorption and scattering efficiencies are also shown.

**Figure 5:** Photocurrent enhancement for bilayer structure having film thicknesses between 1 and 200 nm where the first layer of thickness $d_1$, shown in the inset is either Si or Ge layer whereas the top thin layer of thickness $d_2$ is either $Ge_{0.88}Sn_{0.12}$, $Ge_{0.84}Si_{0.04}Sn_{0.12}$ or $Ge_{0.84}Si_{0.12}Sn_{0.04}$. We present here the short-current enhancement when $d_1 = d_2$.

**Figure 6:** **(a)** 2D short-circuit photocurrent enhancement $\eta_{J_{sc}}$ map as a function of the shell thickness $t$ and the core radius $R_c$ for the CSNW structures: from top-left to bottom-right: Si/$Ge_{0.88}Sn_{0.12}$, Si/ $Ge_{0.84}Si_{0.04}Sn_{0.12}$, Ge/ $Ge_{0.88}Sn_{0.12}$ and Ge/ $Ge_{0.84}Si_{0.04}Sn_{0.12}$. $\eta_{J_{sc}}$ is equal to the absorption efficiency of the CSNW structure divided by the one for the core NW with a radius of $R_c \times (1 + t)$. The horizontal dashed lines represent 3 different shell thicknesses (line a→$t$=3 nm; line b→$t$=50 nm and line c→$t$=100 nm) that will be analyzed more in detail in Fig. 7. Additionally,

we present in panel **(b)** a line profile of Si/ $Ge_{0.88}Sn_{0.12}$. CSNW 2D map, along two shell thickness ($t = [0.25, 1] \times R_c$) directions, shown as solid red and dashed-red lines respectively in panel (a). The red lines represent intensity profiles extracted from the 2D map in panel (a) following the directions $t = R_c$ and $t = R_c/4$. We also present the short-current enhancement of thin-film structure for 2 different top-layer thickness ($d_2 = 0.25d_1$ and $d_2 = d_1$) for a Si/ $Ge_{0.88}Sn_{0.12}$ stack. The red arrows represent the peak core radius positions attributed to LMR.

**Figure 7:** Integrated solar absorption $J_{sc}$ (mA/cm$^2$) as a function of the core radius of the CSNWs consisting of absorbing group IV binary and ternary alloy semiconductors materials. We fix the shell thickness to **(a)** 3 nm, **(b)** 50 nm and **(b)** 100 nm and plot the solar absorption as a function of the core radius for Si/ $Ge_{0.88}Sn_{0.12}$ (*SS1*), Si/ $Ge_{0.84}Si_{0.04}Sn_{0.12}$ (*SS2*), Ge/ $Ge_{0.88}Sn_{0.12}$ (*SS3*), Ge/ $Ge_{0.84}Si_{0.04}Sn_{0.12}$ (*SS4*), GeNW and SiNW structures. The GeNW and SiNW solar absorption are presented for comparison sake, with a core radius equal to $R_c \times (1 + t)$ for a fair comparison, in order to easily visualize the enhancement of light absorption in these structures. We also show the near electric field profile at the highest achievable short-current at a shell thickness of 3 nm and a core radius of 7 nm. We can infer from the profile distribution that the leaky fundamental mode is responsible for such a high short-current.

**Figure 8:** Light absorption enhancement theoretical maps as a function of the core radius and the incident light wavelength for the Si/$Ge_{0.88}Sn_{0.12}$ core-shell nanowire for different shell thicknesses $t = [0.25, 0.5, 0.75, 1] \times R_c$.

**Figure 9: (a)** The short-current enhancement map of the (Si/$Ge_{0.88}Sn_{0.12}$/Si$_3$N$_4$) system where we fixed the inner-shell thickness to be equal to the core radius ($t_{i.s.} = R_c$) and we varied the dielectric capping layer thickness $D$ from 1 to 200 nm. The top panel shows the relative maximum change of the short-current enhancement (max($\eta_{Jsc}$) in %) *vs.* the core radius, whereas the left panel represents the relative change of max($\eta_{Jsc}$) *vs.D*. The relative change is evaluated using the following equation: $(\eta_{Jsc} - 1) \times 100$. The short-current enhancement was evaluated as the ratio of the short-current of the core-multishell nanowire to the short-current of the base CSNW (Si/GeSn) ($\eta_{Jsc} = J_{sc}^{Si/GeSn/SiN} / J_{sc}^{Si/GeSn}$). The orange and green arrows in the top panel represent, respectively, the core radii $R_c$ equal to 13.6 and 78.2 nm, where $J_{sc}$ is enhanced. Next, fixing $R_c$ to the previous radii, we present in panel **(b)** and **(c)**, a 2D map of the absorption efficiency $Q_{abs}$ as a function of the incident wavelength $\lambda$ and the dielectric thickness $D$.

**Figure 10: (a)** TM-like mode absorption efficiency ($Q_{abs}^{TM}$) of the Si NW structure (red curve), Si/ $Ge_{0.88}Sn_{0.12}$ CSNW structure (blue curve) and Si/ $Ge_{0.88}Sn_{0.12}$/ Si$_3$N$_4$ structure (green curve), surrounded by air, for 2 different core radii: $R_c = 13.6$ and 78 nm. We kept the GeSn shell thickness fixed to the core radius and the dielectric capping layer $D$ is chosen to be 33 and 45 nm, respectively for each core radius. We also label the resonant peaks 1-10 for the different structures. **(b)** Near field magnitude for the total TM-polarized Poynting vector $|\mathbf{S}|_{TM}^2$ by the three different structures evaluated with the analytical solution for a perpendicular incident illumination at the wavelengths corresponding to the peaks labeled 1−10 in panel (A). We also present the corresponding scale bar for each structure.

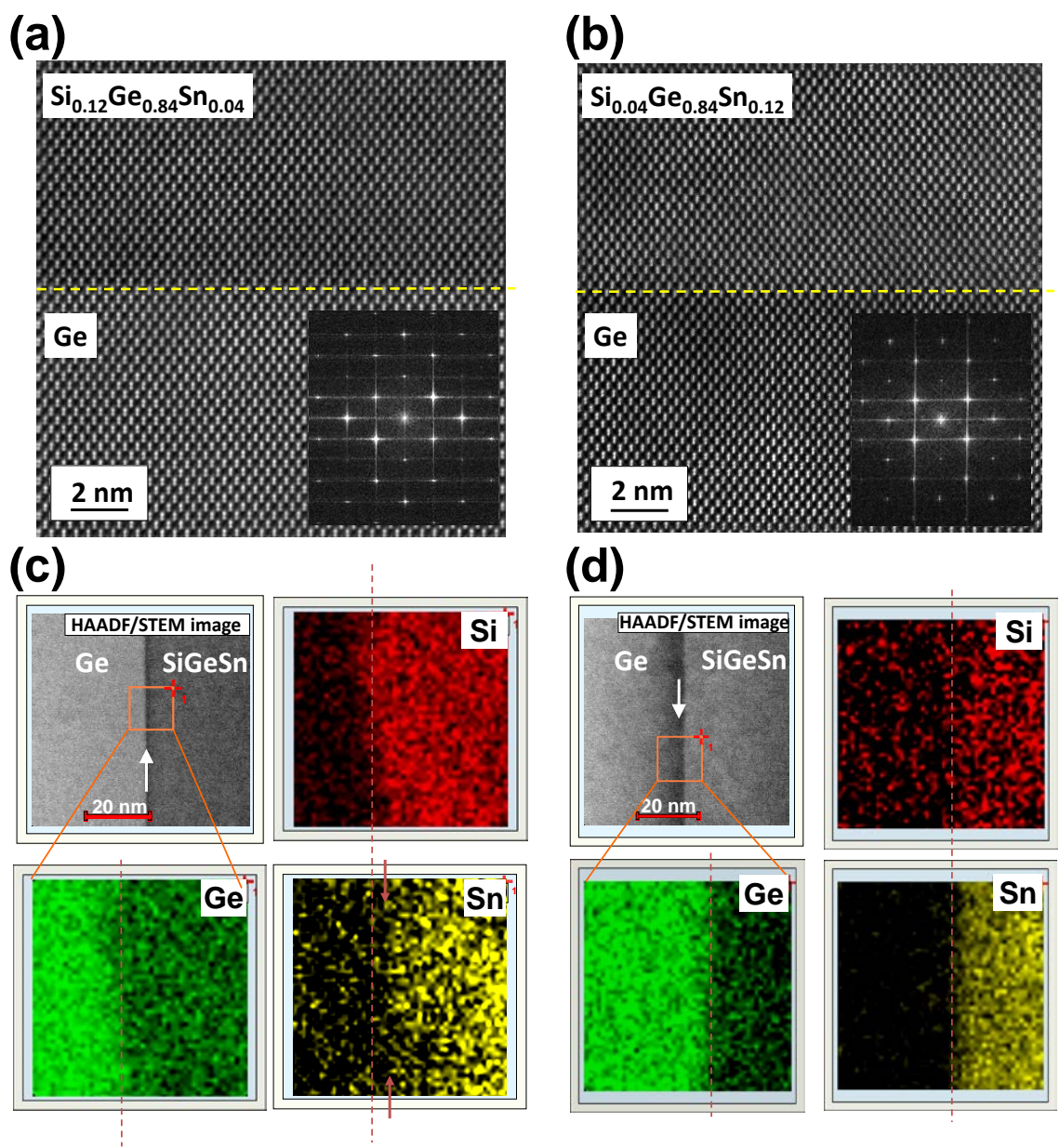

**Figure 1:** High Angle Annular Dark Field Scanning Transmission Electron Microscopy (HAADF/STEM) image of $Ge_{0.84}Si_{0.12}Sn_{0.04}$ **(a)** $Ge_{0.84}Si_{0.04}Sn_{0.12}$ **(b)** layers grown on Ge virtual substrates. Note the absence of dislocations or extended defects in the ternary layer or at the interface. The corresponding diffraction patterns measured at the interface are shown as inset figures in (a) and (b) confirming the high crystallinity as well as the absence of Sn precipitates. Low magnification HAADF/STEM images and EDX maps of Si, Ge, and Sn in $Ge_{0.84}Si_{0.12}Sn_{0.04}$ **(c)** and $Ge_{0.84}Si_{0.04}Sn_{0.12}$ **(d)** layers.

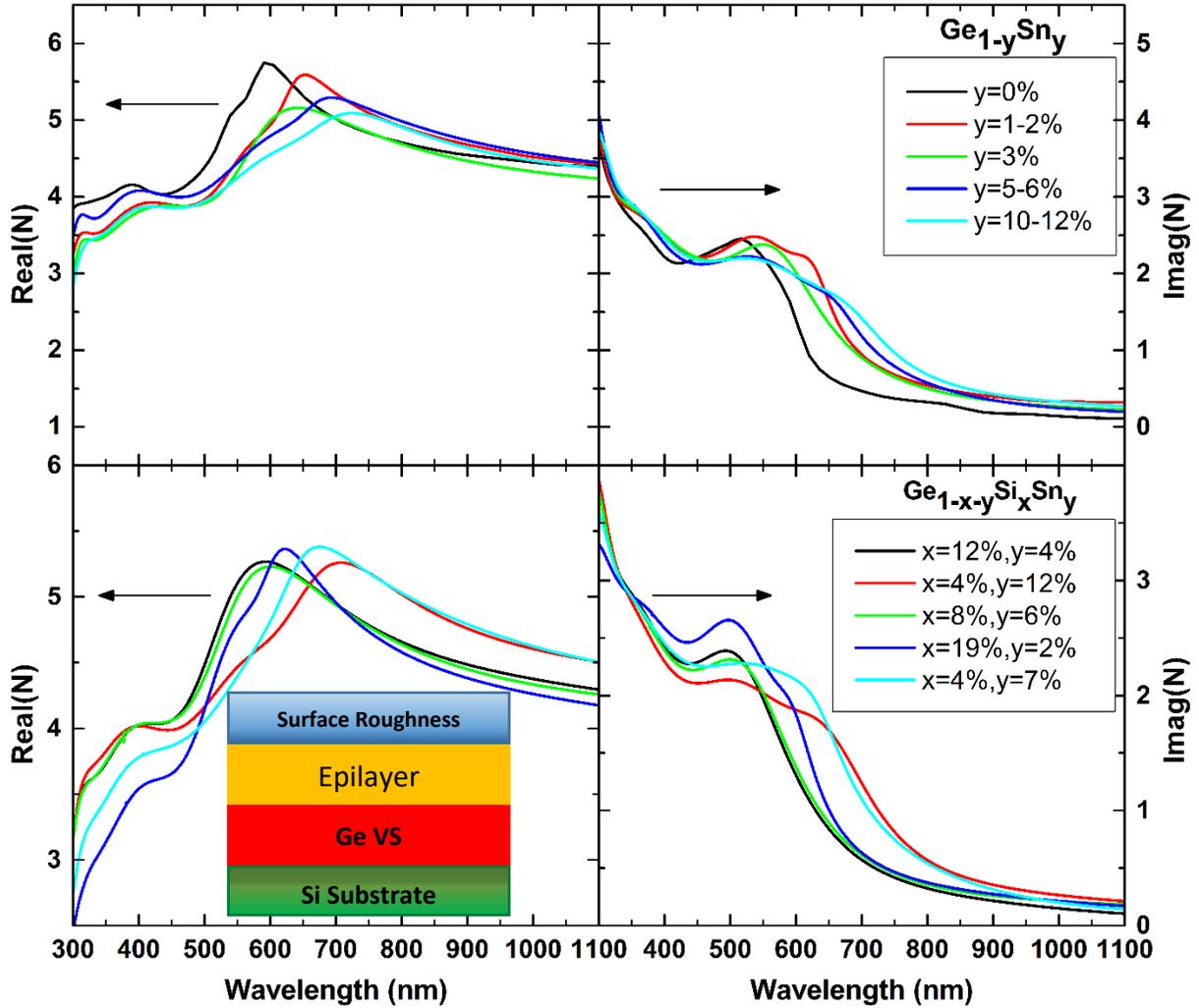

**Figure 2:** The complex refractive index constant $N(\lambda)$ spectra of (top) $Ge_{1-x}Sn_x$ binary semiconductor alloy as a function of Sn composition and (bottom) of $Ge_{1-x-y}Si_xSn_y$ ternary alloy as function of Si and Sn composition obtained from spectroscopic ellipsometry measurement. The optical properties of Ge (001) substrate (x=0%) in the $Ge_{1-x}Sn_x$ binary are taken from Palik [58]. The inset shows the multilayer model used to extract the optical properties (details in SI, section S1).

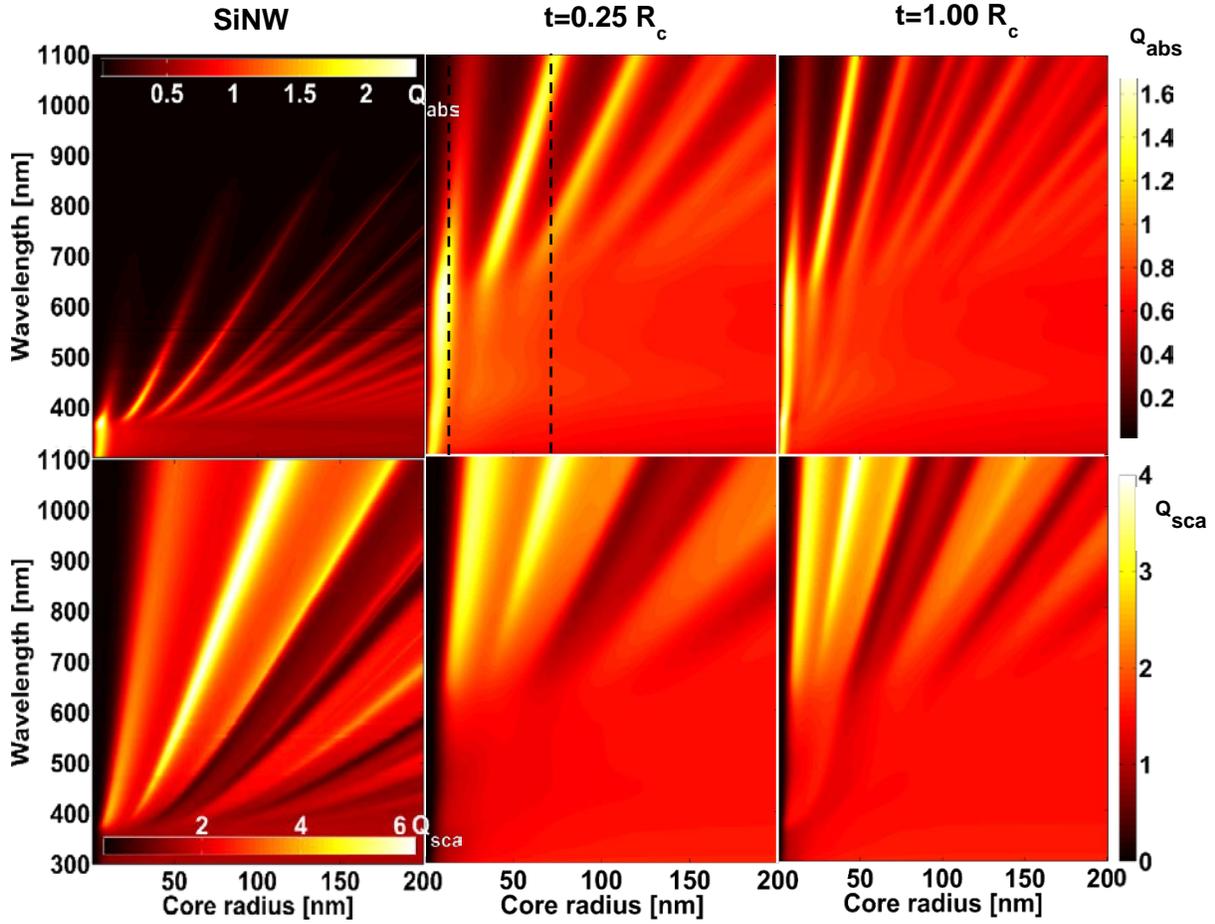

**Figure 3:** Unpolarized absorption efficiency $Q_{abs}$ (top) and unpolarized scattering efficiency $Q_{sca}$ (bottom) of Si/Ge$_{0.88}$Sn$_{0.12}$ CSNW surrounded by air as a function of the core radius $R_c$ and the incident light wavelength for two different shell thicknesses: $t = R_c/4$ and $t = R_c$. For comparison, SiNW absorption and scattering efficiencies are also shown. The three black-dashed lines indicate the selected radii: on-resonance (at $R_c = 8\ nm$) and off-resonance (at $R_c = 60\ nm$).

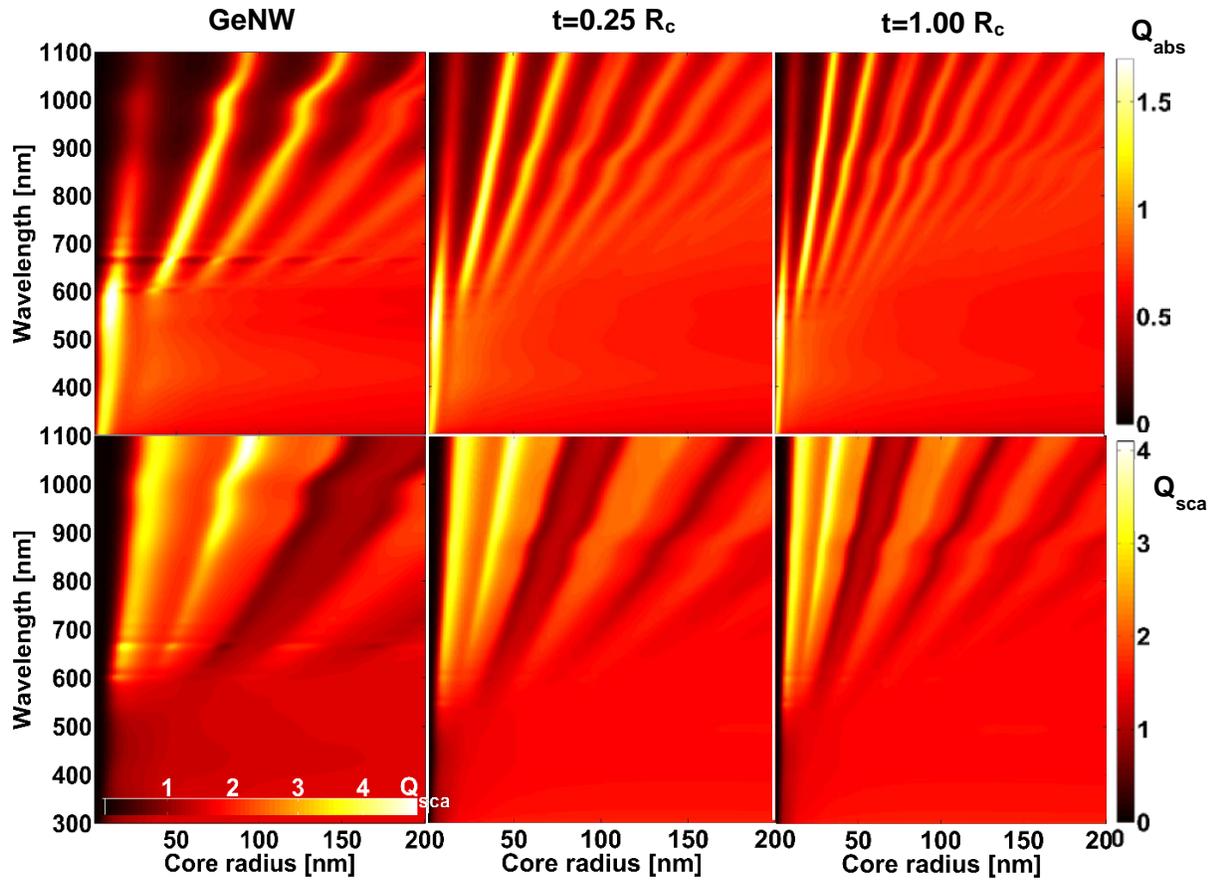

**Figure 4:** Unpolarized absorption efficiency $Q_{abs}$ (top) and unpolarized scattering efficiency $Q_{sca}$ (bottom) of Ge/ $Ge_{0.88}Sn_{0.12}$ CSNW surrounded by air as a function of the core radius $R_c$ and the incident light wavelength for two different shell thicknesses: $t = R_c/4$ and $t = R_c$. For comparison, GeNW absorption and scattering efficiencies are also shown.

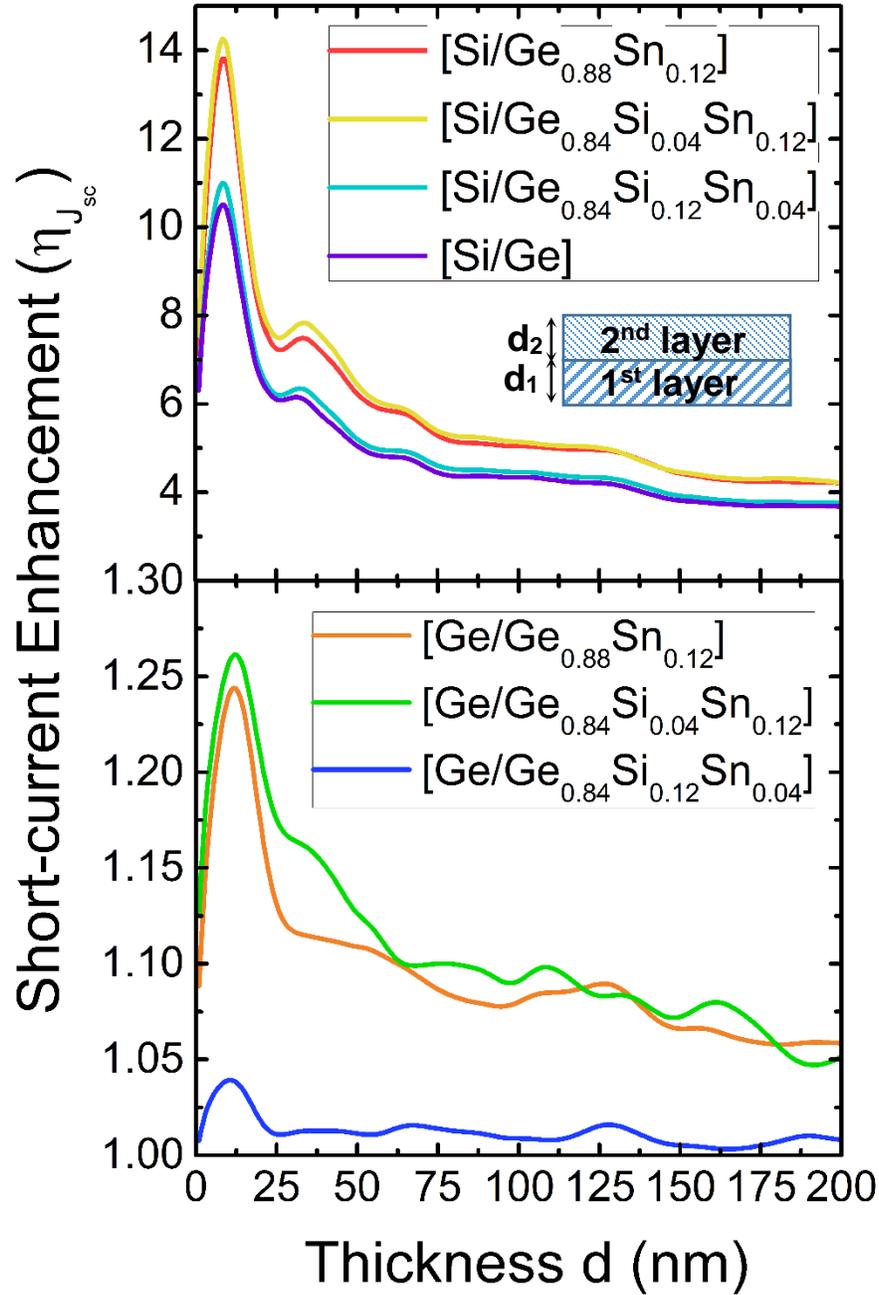

**Figure 5:** Photocurrent enhancement for bilayer structure having film thicknesses between 1 and 200 nm where the first layer of thickness $d_1$, shown in the inset is either Si or Ge layer whereas the top thin layer of thickness $d_2$ is either $Ge_{0.88}Sn_{0.12}$, $Ge_{0.84}Si_{0.04}Sn_{0.12}$ or $Ge_{0.84}Si_{0.12}Sn_{0.04}$. We present here the short-current enhancement when $d_1 = d_2$.

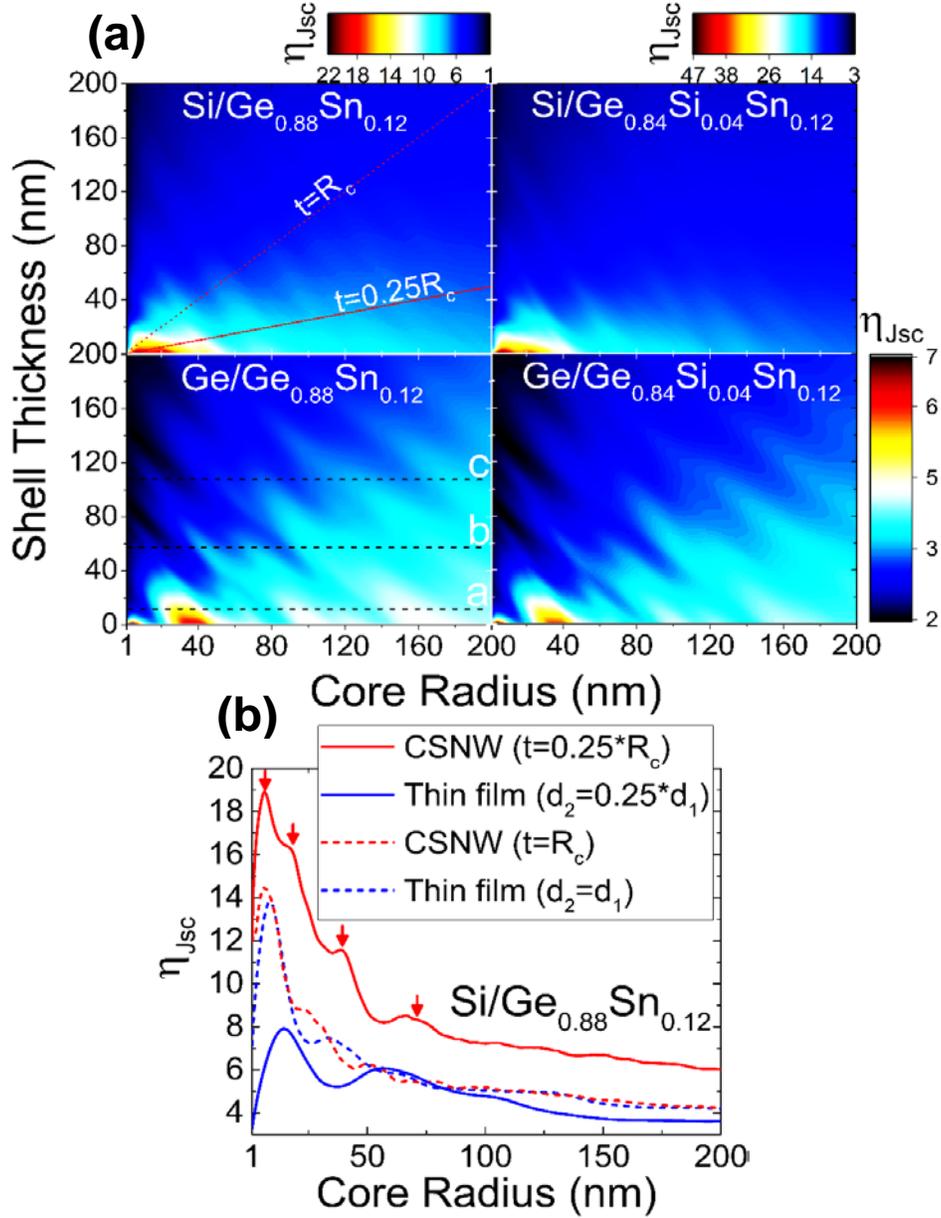

**Figure 6:** **(a)** 2D short-circuit photocurrent enhancement $\eta_{Jsc}$ map as a function of the shell thickness $t$ and the core radius $R_c$ for the CSNW structures: from top-left to bottom-right: Si/Ge$_{0.88}$Sn$_{0.12}$, Si/ Ge$_{0.84}$Si$_{0.04}$Sn$_{0.12}$, Ge/ Ge$_{0.88}$Sn$_{0.12}$ and Ge/ Ge$_{0.84}$Si$_{0.04}$Sn$_{0.12}$. $\eta_{Jsc}$ is equal to the absorption efficiency of the CSNW structure divided by the one for the core NW with a radius of $R_c \times (1+t)$. The horizontal dashed lines represent 3 different shell thicknesses (line a→$t$=3 nm; line b→$t$=50 nm and line c→$t$=100 nm) that will be analyzed more in detail in Fig. 7. Additionally, we present in panel **(b)** a line profile of Si/ Ge$_{0.88}$Sn$_{0.12}$. CSNW 2D map, along two shell thickness ($t = [0.25, 1] \times R_c$) directions, shown as solid red and dashed-red lines respectively in panel (a). The red lines represent intensity profiles extracted from the 2D map in panel a following the directions $t = R_c$ and $t = R_c/4$. We also present the short-current enhancement of thin-film structure for 2 different top-layer thickness ($d_2 = 0.25d_1$ and $d_2 = d_1$) for a Si/ Ge$_{0.88}$Sn$_{0.12}$ stack. The red arrows represent the peak core radius positions attributed to LMR.

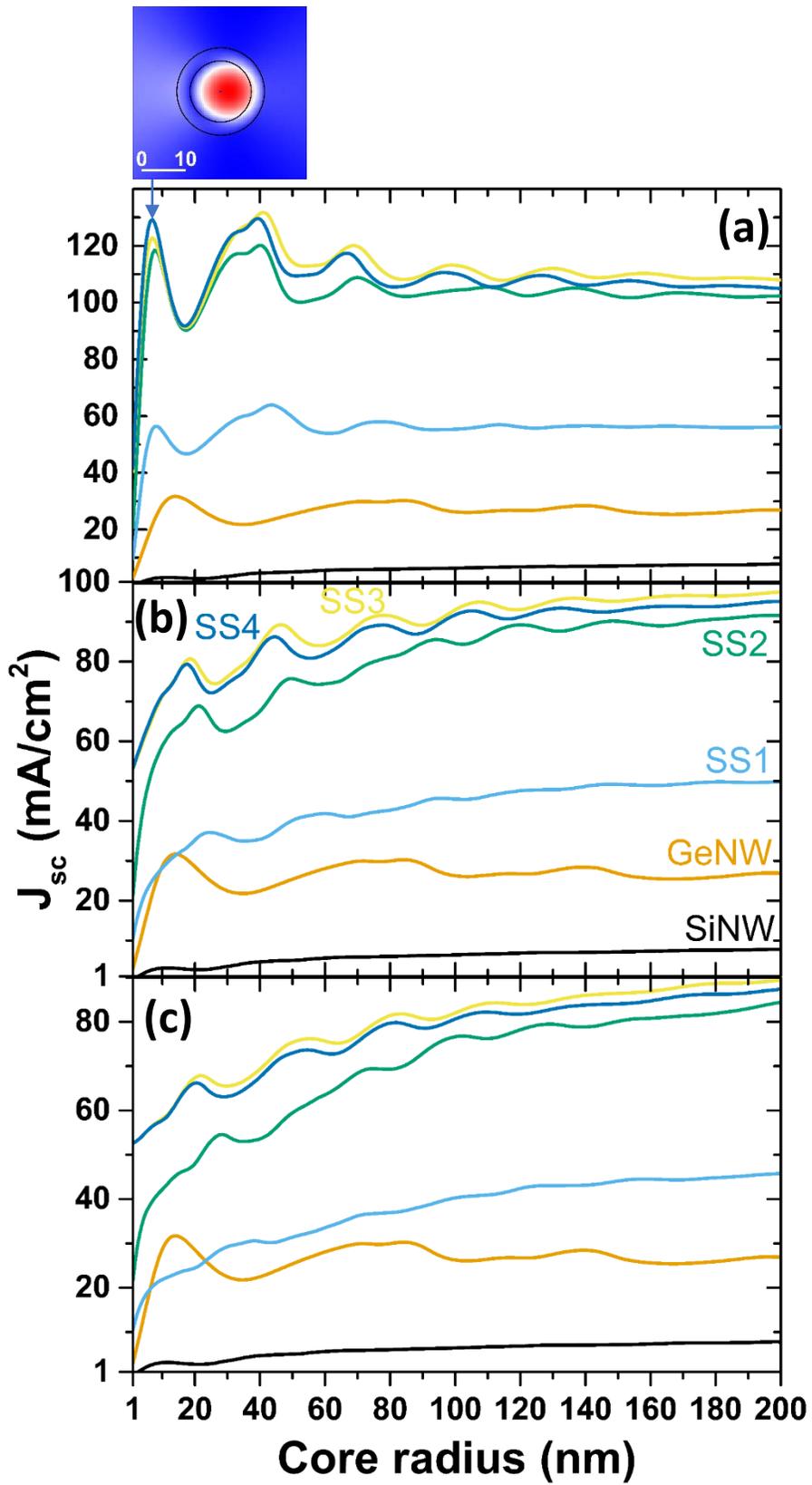

**Figure 7:** Integrated solar absorption $J_{sc}$ (mA/cm$^2$) as a function of the core radius of the CSNWs consisting of absorbing group IV binary and ternary alloy semiconductors materials. We fix the shell thickness to **(a)** 3 nm, **(b)** 50 nm and **(b)** 100 nm and plot the solar absorption as a function of the core radius for Si/ Ge$_{0.88}$Sn$_{0.12}$ (*SS1*), Si/ Ge$_{0.84}$Si$_{0.04}$Sn$_{0.12}$ (*SS2*), Ge/ Ge$_{0.88}$Sn$_{0.12}$ (*SS3*), Ge/ Ge$_{0.84}$Si$_{0.04}$Sn$_{0.12}$ (*SS4*), GeNW and SiNW structures. The GeNW and SiNW solar absorption are presented for comparison sake, with a core radius equal to $R_c \times (1 + t)$ for a fair comparison, in order to easily visualize the enhancement of light absorption in these structures. We also show the near electric field profile at the highest achievable short-current at a shell thickness of 3 nm and a core radius of 7 nm. We can infer from the profile distribution that the leaky fundamental mode is responsible for such a high short-current.

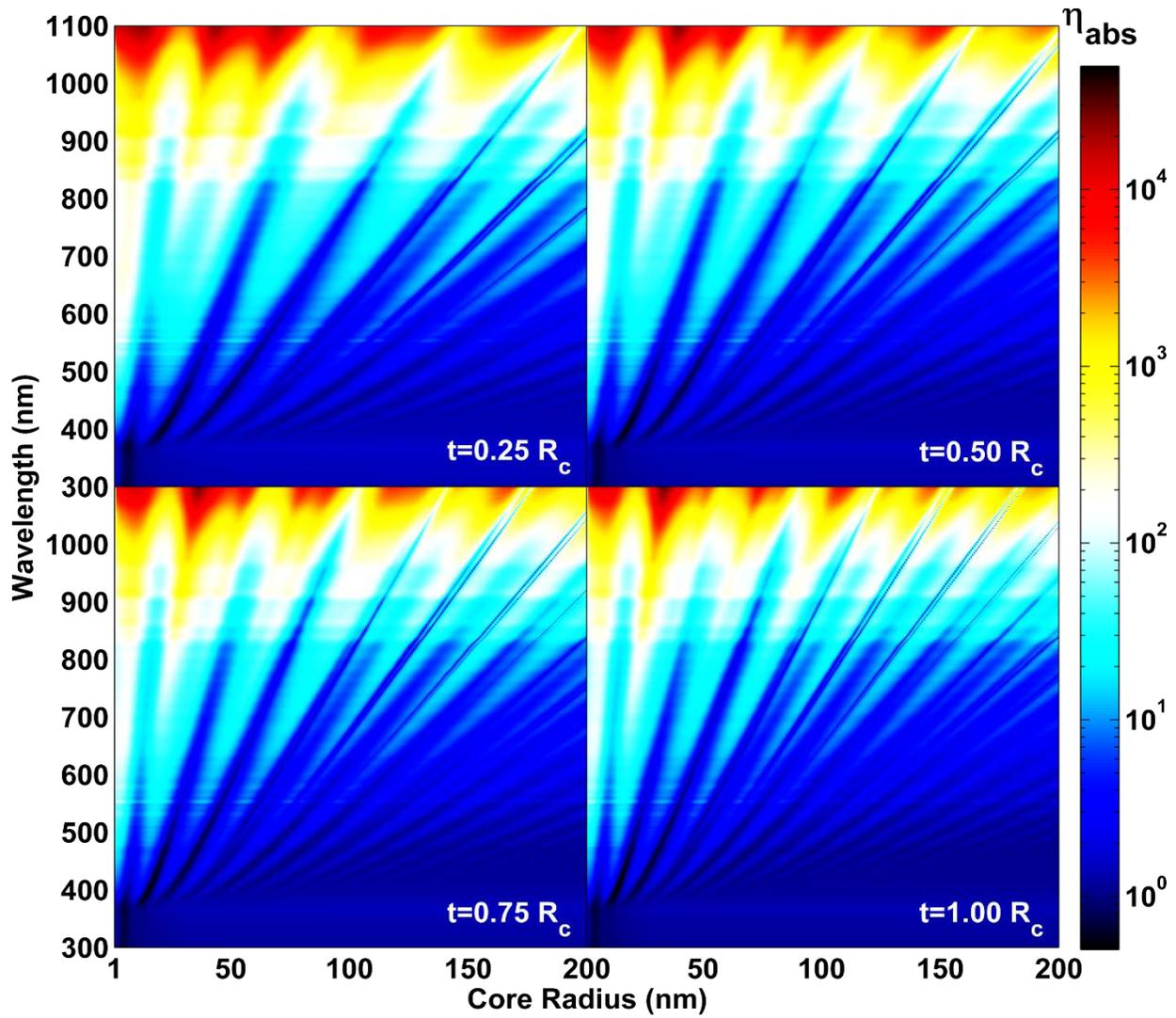

**Figure 8:** Extreme light absorption enhancement map as a function of the core radius and the incident light wavelength for the Si/Ge$_{0.88}$Sn$_{0.12}$ core-shell nanowire for different shell thicknesses $t = [0.25, 0.5, 0.75, 1] \times R_c$.

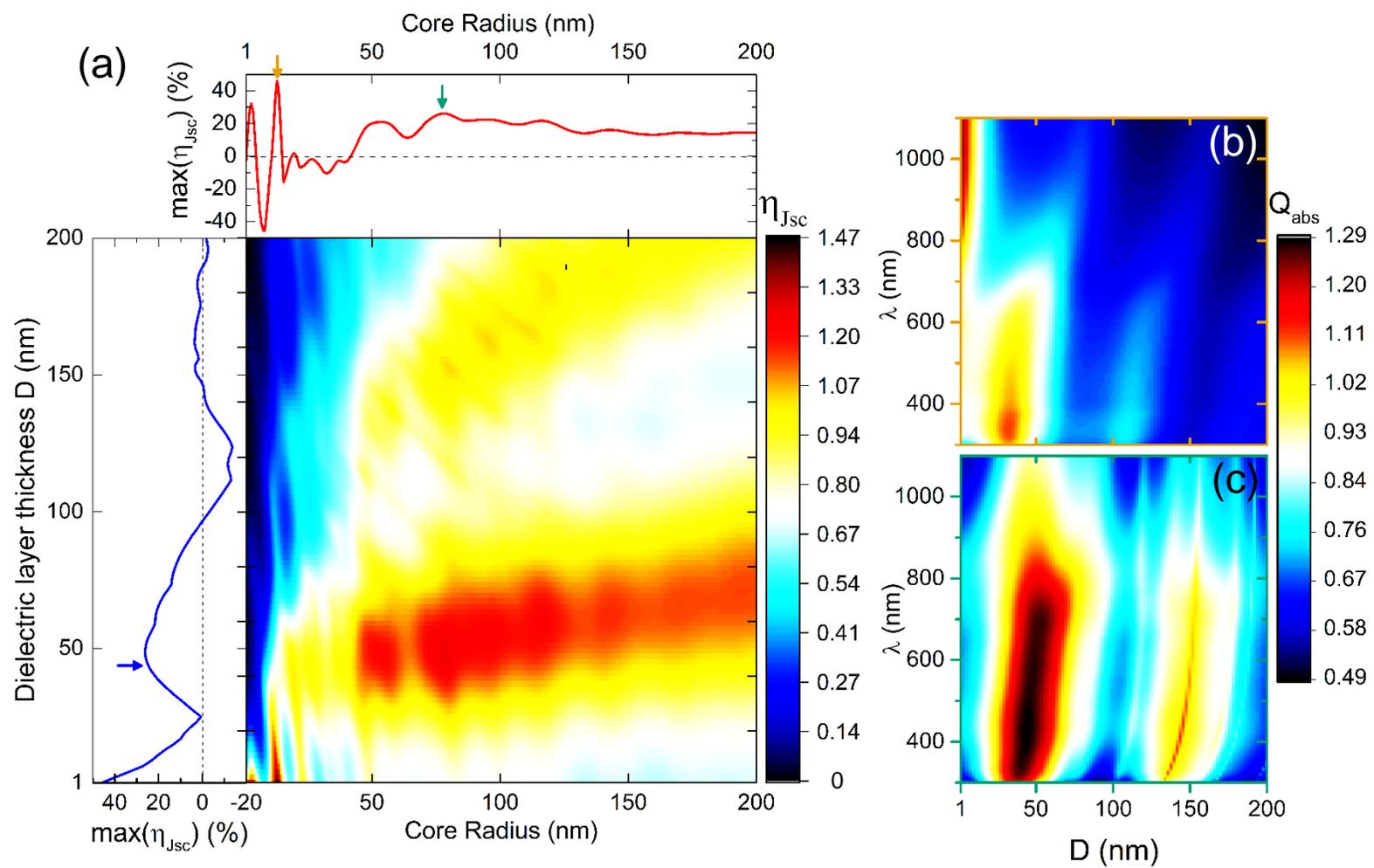

**Figure 9: (a)** The short-current enhancement map of the (Si/Ge$_{0.88}$Sn$_{0.12}$/Si$_3$N$_4$) system where we fixed the inner-shell thickness to be equal to the core radius ($t_{i.s.} = R_c$) and we varied the dielectric capping layer thickness $D$ from 1 to 200 nm. The top panel shows the relative maximum change of the short-current enhancement (max$(\eta_{Jsc})$ in %) *vs.* the core radius, whereas the left panel represents the relative change of max$(\eta_{Jsc})$ *vs.* $D$. The relative change is evaluated using the following equation: $(\eta_{Jsc} - 1) \times 100$. The short-current enhancement was evaluated as the ratio of the short-current of the core-multishell nanowire to the short-current of the base CSNW (Si/GeSn) ($\eta_{Jsc} = J_{sc}^{\text{Si/GeSn/SiN}} / J_{sc}^{Si/GeSn}$). The orange and green arrows in the top panel represent, respectively, the core radii $R_c$ equal to 13.6 and 78.2 nm, where $J_{sc}$ is enhanced. Next, fixing $R_c$ to the previous radii, we present in panel **(b)** and **(c)**, a 2D map of the absorption efficiency $Q_{abs}$ as a function of the incident wavelength $\lambda$ and the dielectric thickness $D$.

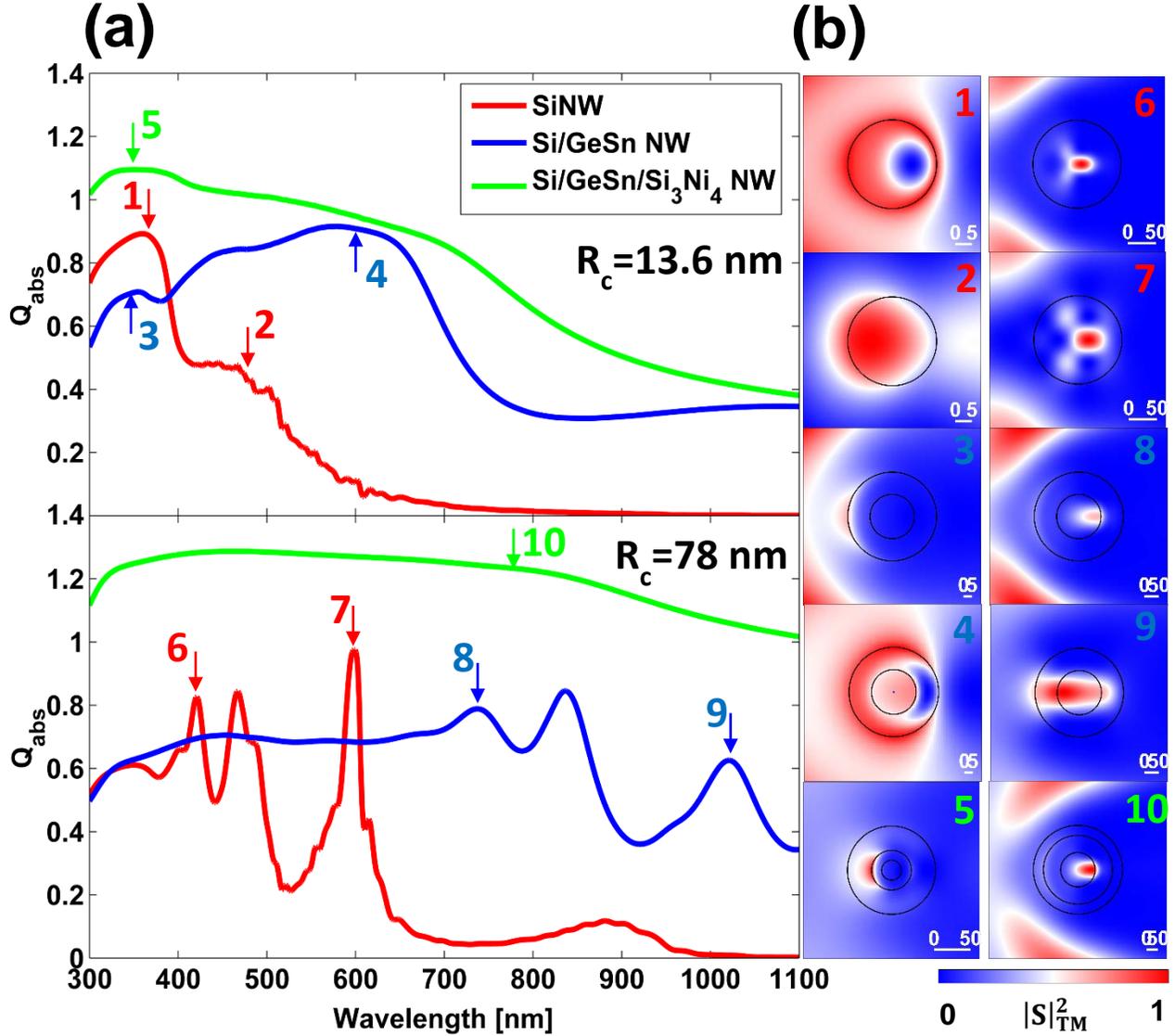

**Figure 10: (a)** TM-like mode absorption efficiency ($Q_{abs}^{TM}$) of the Si NW structure (red curve), Si/ Ge$_{0.88}$Sn$_{0.12}$ CSNW structure (blue curve) and Si/ Ge$_{0.88}$Sn$_{0.12}$/ Si$_3$N$_4$ structure (green curve), surrounded by air, for 2 different core radii: $R_c = 13.6$ and 78 nm. We kept the GeSn shell thickness fixed to the core radius and the dielectric capping layer $D$ is chosen to be 33 and 45 nm, respectively for each core radius. We also label the resonant peaks 1-10 for the different structures. **(b)** Near field magnitude for the total TM-polarized Poynting vector $|\mathbf{S}|_{TM}^2$ by the three different structures evaluated with the analytical solution for a perpendicular incident illumination at the wavelengths corresponding to the peaks labeled 1−10 in panel (A). We also present the corresponding scale bar for each structure.

# Supporting Information

# Enhanced IR Light Absorption in Group IV-SiGeSn Core-Shell Nanowires

Anis Attiaoui, [1] Stephan Wirth, [2] André-Pierre Blanchard-Dionne, [1] Michel Meunier, [1] J. M. Hartmann, [3] Dan Buca, [2] and Oussama Moutanabbir [1, *]

[1]*Department of Engineering Physics, École Polytechnique de Montréal, Montréal, C.P. 6079, Succ. Centre-Ville, Québec, H3C 3A7 Canada*

[2]*Peter Grünberg Institute 9 and JARA-FIT, Forschungszentrum Juelich, 52425 Juelich, Germany*

[3]*CEA, LETI, Minatec Campus, 17 rue des Martyrs, 38054 Grenoble, France*

# S1. Optical characterisation of GeSn and GeSiSn alloy

The optical measurements of the studied systems were performed with a commercial ellipsometer (variable-angle spectroscopic ellipsometer J. A. Woollam).[1] However, a fundamental issue for ellipsometric measurement is to accurately analyze and model the desired layer due to the formation of a native oxide on GeSn ($GeO_2$) and GeSiSn ($SiO_2$) layers, which cannot be prevented despite of the etch. Therefore, it is detrimental to include surface oxide layers when building the ellipsometric model in order to extract the optical constants from the complex ellipsometric ratio $\rho$, which is the ratio of the reflectance coefficients $r_p$ and $r_s$ (parallel and perpendicular to the plane of incidence). This is generally expressed in terms of two angles $\Psi$ and $\Delta$:

$$\rho = \frac{r_p}{r_s} = e^{i\Delta} \tan \Psi \qquad (1)$$

where $\Psi$ is a measure of the relative amplitude and $\Delta$ the relative shift. From $\Psi$ and $\Delta$, the complex pseudo-dielectric function $\hat{\varepsilon}(\lambda) = \varepsilon_1(\lambda) + i\varepsilon_2(\lambda)$ or equivalently the complex refractive index $\boldsymbol{N} = n + ik$ can be readily derived using a two-phase model.[2]

The pseudo-dielectric function is not the true dielectric functions of the investigated layers as it includes the effect of over-layers and the underlying substrates (Si and Ge). Nevertheless, it allows an acceptable interpretation of the dielectric response of the sample. The ellipsometric data were analyzed in term of a four-layer model- as shown in the inset of Figure 2- consisting of a Si substrate, a Ge virtual substrate, a $Ge_{1-y}Sn_y$ or $Ge_{1-x-y}Si_xSn_y$ alloy, and an oxide surface layer. The dielectric function of $Ge_{1-y}Sn_y$ was analyzed with a "parametric optical constant model" developed by Johs and Herzinger.[3,4] The adjustable parameters of our model are the surface layer thickness, the $Ge_{1-y}Sn_y$ or $Ge_{1-x-y}Si_xSn_y$ thickness, the Ge (VS) thickness and the Johs-Herzinger model that describes the dielectric function of the investigated alloys. They were investigated using a proprietary Marquardt-Levenberg algorithm provided by Woollam ellipsometer's manufacturer.

The known dielectric function of Si substrate as well as that of GeO$_2$ [5] were used in their tabulated form. Initially, the data from Palik[6] was used for Ge virtual substrate. However, in order to enhance the optical model, we decided later on to etch the GeSn layer from a different reference sample with the highest Sn composition (Si/Ge (VS)/Ge$_{0.88}$Sn$_{0.12}$) in order to obtain only the bare Ge virtual substrate, which we would then optically characterize in order to extract the optical properties of the Ge VS used during the RPCVD growth of the samples. We used a chemical wet etching procedure, a CYANTEK solution composed of mixture of different solutions (H$_3$PO$_4$:C$_2$H$_4$O$_2$:HN$_3$: H$_2$O) with a volume weight ratio of 72:3:3:22. To verify the quality of the etching, multi-wavelength micro-Raman spectroscopy was employed to confirm the complete etching of the GeSn layer. After investigating the optical properties of the etched surface using VASE, we extracted the ellipsometric parameters (Ψ and Δ) and we saved the obtained model and incorporated it when analyzing the studied samples presented in Figure 2 in the main text.

## S2.    Lorentz-Mie scattering theory for an infinite core-shell cylindrical nanowire.

In order to quantify, the absorption and scattering efficiencies in the core-shell nanowire, we need to know the corresponding absorption and scattering expansion coefficients. Furthermore, we need to quantify the electrical $\vec{E}$ and magnetic fields $\vec{H}$ inside the core and the shell in addition to the incident and scattering fields. We present in Figure S1 a schematic representation for the studied core-shell cylindrical nanowire under the Mie scattering formalism.

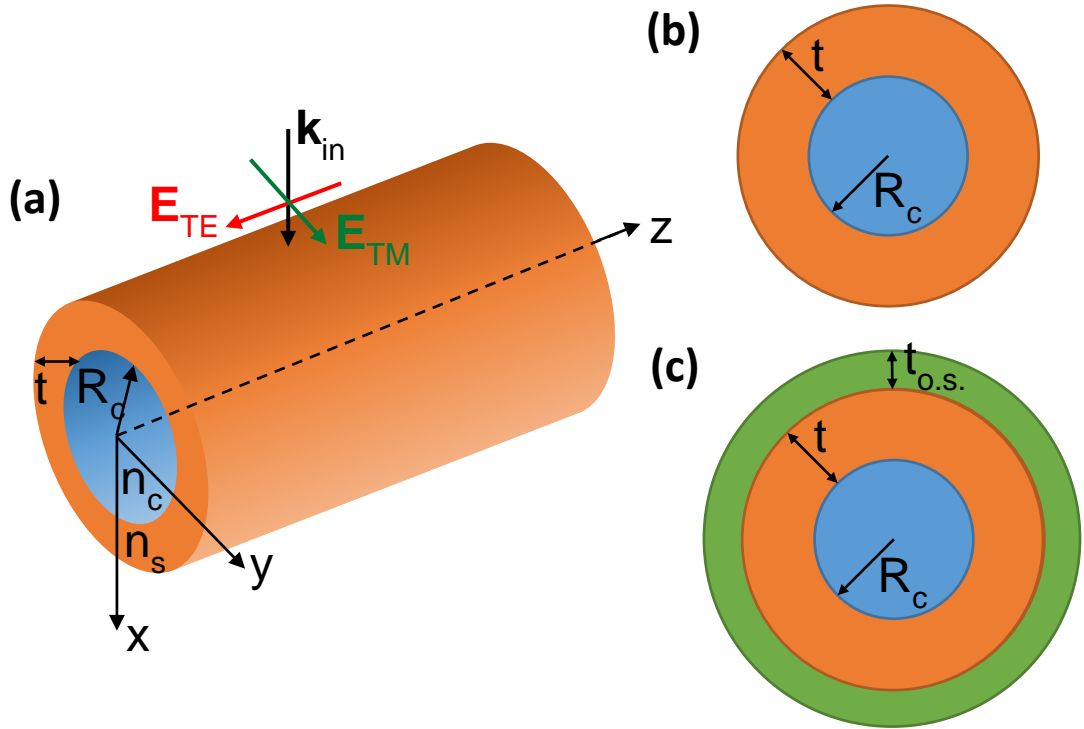

**Supplementary Figure S1**: **(a)** Schematic of the analytical solution of light interaction with an infinite cylindrical core-shell nanowire. The core and the shell are both formed with group IV semiconductor binary or ternary alloy having respectively different complex refractive index $n_c$ and $n_s$. Cross-section of a cylindrical **(b)** CSNW system with a core radius $R_c$ and a shell thickness $t$ and a cylindrical **(c)** core-multishell nanowire system having a core radius $R_c$, an inner-shell of thickness $t$ and an outer-shell thickness $t_{o.s.}$ with the respective complex refractive indices $n_c, n_{s,}$ and $n_{o.s.}$. The material composing the outer-shell in panel (c) will either be a non-absorbing dielectric or GeSiSn semiconductor.

In order to include the effect of the incidence angle in our calculation, we present a general expression of the electromagnetic fields in the CSNW. For that, lets us consider a long cylindrical core-shell nanowire as shown in Figure S1 (where the length $L$ is almost 10 times larger than the radial dimensions $r$) oriented along the $\boldsymbol{e}_z$ direction and having a core of radius $r_c$ and a complex refractive index $N_c$ and a shell of radius $r_s$ with a complex refractive index $N_s$. It is better to define some reduced constant in order to simplify the final expression of the electromagnetic fields and thus the scattering and absorption efficiencies. In each medium, the wavenumber is given by $k_j = 2\pi N_j/\lambda$ where $j \in \{\boldsymbol{M}, \boldsymbol{c}, \boldsymbol{s}\}$ for an arbitrary nonabsorbing

*M*edium in which the CSNW is embedded, the *c*ore and the *s*hell respectively and $N_j$ is the corresponding complex refractive index. Besides, we define an incidence angle $\xi$ between the x- and z- axis. Finally, we define a dimensionless radius $\rho_j(r) = r\sqrt{k_j^2 - h^2}$ where $h_j = -k_j \cos\xi$ and $j \in \{M, c, s\}$ will be used. Furthermore, $r$ can take either value of the outer radius $r_s$ or the inner radius $r_c$.

We start by defining the general electromagnetic (EM) field's expression in three regions: the outer region of the nanowire where are the incident $\{E_{inc}, H_{inc}\}$ and scattered $\{E_{sca}, H_{sca}\}$ EM fields, the shell region $\{E_s, H_s\}$, and the core region $\{E_c, H_c\}$. The total EM field outside the nanowire $\{E_t, H_t\}$ is the sum of both incident and scattered fields. Extending the work of Ref. [7,8], we present a general expression for the unpolarized EM fields, then the TE and TM modes while considering an arbitrary angle of incidence. Firstly, the incident EM field is given by

$$\boldsymbol{E}_{inc} = \frac{E_0}{k_M \sin\xi} \sum_{n=-\infty}^{\infty} (-i)^n \left\{ \boldsymbol{N}_n^{(1)}(\rho_M) - i\boldsymbol{M}_n^{(1)}(\rho_M) \right\} e^{in\phi} e^{ihz}$$

$$\boldsymbol{H}_{inc} = \frac{-ik_M}{\omega\,\mu_0} \frac{E_0}{k_M \sin\xi} \sum_{n=-\infty}^{\infty} (-i)^n \left\{ \boldsymbol{N}_n^{(1)}(\rho_M) - i\boldsymbol{M}_n^{(1)}(\rho_M) \right\} e^{in\phi} e^{ihz} \tag{2}$$

Where $\boldsymbol{N}_n^{(1)}(\rho_M) = \frac{\sqrt{k_M^2 - h_M^2}}{k_M} \begin{bmatrix} ih_M J_n'(\rho_M) \\ -h_M n J_n(\rho_M)/\rho_M \\ \sqrt{k_M^2 - h_M^2} J_n(\rho_M) \end{bmatrix}$, $\boldsymbol{M}_n^{(1)}(\rho_M) = \sqrt{k_M^2 - h_M^2} \begin{bmatrix} in J_n(\rho_M)/\rho_M \\ J_n'(\rho_M) \\ 0 \end{bmatrix}$

where $\omega$ and $\mu_0$ are the angular frequency and the permeability of free space, respectively.

Secondly, the EM fields of the scattered wave are:

$$\boldsymbol{E}_{sca} = \frac{E_0}{k_M \sin\xi} \sum_{n=-\infty}^{\infty} (-i)^n \left\{ a_n \boldsymbol{N}_n^{(3)}(\rho_M) - i b_n \boldsymbol{M}_n^{(3)}(\rho_M) \right\} e^{in\phi} e^{ihz}$$

$$\boldsymbol{H}_{sca} = \frac{-ik_M}{\omega\,\mu_0} \frac{E_0}{k_M \sin\xi} \sum_{n=-\infty}^{\infty} (-i)^n \left\{ b_n \boldsymbol{N}_n^{(3)}(\rho_M) - i a_n \boldsymbol{M}_n^{(3)}(\rho_M) \right\} e^{in\phi} e^{ihz} \tag{3}$$

Where $\boldsymbol{N}_n^{(3)}(\rho_M) = \frac{\sqrt{k_M^2 - h_M^2}}{k_M} \begin{bmatrix} ih_M H_n'(\rho_M) \\ -h_M n H_n(\rho_M)/\rho_M \\ \sqrt{k_M^2 - h_M^2} H_n(\rho_M) \end{bmatrix}$, $\boldsymbol{M}_n^{(3)}(\rho_M) = \sqrt{k_M^2 - h_M^2} \begin{bmatrix} in\, H_n(\rho_M)/\rho_M \\ H_n'(\rho_M) \\ 0 \end{bmatrix}$

where $a_n$ and $b_n$ are the scattering field expansion coefficients.

Thirdly, the EM fields of the wave inside the core are:

$$\boldsymbol{E}_c = \frac{E_0}{k_c \sin\xi} \sum_{n=-\infty}^{\infty} (-i)^n \left[ \alpha_n^c \boldsymbol{N}_n^{(1)}(\rho_c) - i\beta_n^c \boldsymbol{M}_n^{(1)}(\rho_c) \right] e^{in\phi} e^{ihz}$$

$$\boldsymbol{H}_c = \frac{-ik_c}{\omega\,\mu_0} \frac{E_0}{k_c \sin\xi} \sum_{n=-\infty}^{\infty} (-i)^n \left[ \beta_n^c \boldsymbol{N}_n^{(1)}(\rho_c) - i\alpha_n^c \boldsymbol{M}_n^{(1)}(\rho_c) \right] e^{in\phi} e^{ihz} \qquad (4)$$

Where $\boldsymbol{N}_n^{(1)}(\rho_c) = \frac{\sqrt{k_c^2 - h_c^2}}{k_c} \begin{bmatrix} ih_c J_n'(\rho_c) \\ -h_c n J_n(\rho_c)/\rho_c \\ \sqrt{k_M^2 - h_c^2} J_n(\rho_c) \end{bmatrix}$, $\boldsymbol{M}_n^{(1)}(\rho_c) = \sqrt{k_c^2 - h_c^2} \begin{bmatrix} in J_n(\rho_c)/\rho_c \\ J_n'(\rho_c) \\ 0 \end{bmatrix}$

Finally, the EM fields of the wave inside the shell are:

$$\boldsymbol{E}_s = \frac{E_0}{k_s \sin\xi} \sum_{n=-\infty}^{\infty} (-i)^n \left\{ \left[ \alpha_n^s \boldsymbol{N}_n^{(1)}(\rho_s) + \gamma_n^s \boldsymbol{N}_n^{(2)}(\rho_s) \right] \right.$$
$$\left. - i \left[ \beta_n^s \boldsymbol{M}_n^{(1)}(\rho_s) + \delta_n^s \boldsymbol{M}_n^{(2)}(\rho_s) \right] \right\} e^{in\phi} e^{ihz}$$

$$\boldsymbol{H}_s = \frac{-ik_s}{\omega\,\mu_0} \frac{E_0}{k_s \sin\xi} \sum_{n=-\infty}^{\infty} (-i)^n \left\{ \left[ \beta_n^s \boldsymbol{N}_n^{(1)}(\rho_s) + \delta_n^s \boldsymbol{N}_n^{(2)}(\rho_s) \right] \right.$$
$$\left. - i \left[ \alpha_n^s \boldsymbol{M}_n^{(1)}(\rho_s) + \gamma_n^s \boldsymbol{M}_n^{(2)}(\rho_s) \right] \right\} e^{in\phi} e^{ihz} \qquad (5)$$

Where $\boldsymbol{N}_n^{(1)}(\rho_s) = \frac{\sqrt{k_s^2 - h_s^2}}{k_s} \begin{bmatrix} ih_s J_n'(\rho_s) \\ -h_s n J_n(\rho_s)/\rho_s \\ \sqrt{k_s^2 - h^2} J_n(\rho_s) \end{bmatrix}$, $\boldsymbol{N}_n^{(2)}(\rho_s) = \frac{\sqrt{k_s^2 - h_s^2}}{k_s} \begin{bmatrix} ih_s Y_n'(\rho_s) \\ -h_s n Y_n(\rho_s)/\rho_s \\ \sqrt{k_s^2 - h_s^2} Y_n(\rho_s) \end{bmatrix}$,

$\boldsymbol{M}_n^{(1)}(\rho_s) = \sqrt{k_s^2 - h_s^2} \begin{bmatrix} in J_n(\rho_s)/\rho_s \\ J_n'(\rho_s) \\ 0 \end{bmatrix}$, $\boldsymbol{M}_n^{(2)} = \sqrt{k_s^2 - h_s^2} \begin{bmatrix} in Y_n(\rho_s)/\rho_s \\ H_n'(\rho_s) \\ 0 \end{bmatrix}$

where $\alpha_n^j$, $\gamma_n^j$ are the expansion coefficients of the TM modes and $\beta_n^j$, $\delta_n^j$ are those of the TE modes in the shell ($j = s$) and in the core ($j = c$) and $a_n$ and $b_n$ are the expansion coefficients of the TM and TE modes, respectively, of the scattered wave.[9]

Next, Applying Maxwell's boundary conditions at the interfaces and the cylinder surface, the expansion coefficients of all waves inside the multilayered cylinder and of the scattered wave can be resolved by evaluating the following continuity equations

$$(\boldsymbol{E}_s - \boldsymbol{E}_c) \times \boldsymbol{e}_r = \boldsymbol{0} \; ; \; (\boldsymbol{H}_s - \boldsymbol{H}_c) \times \boldsymbol{e}_r = \boldsymbol{0}$$

$$(\boldsymbol{E}_{sca} + \boldsymbol{E}_{inc} - \boldsymbol{E}_s) \times \boldsymbol{e}_r = \boldsymbol{0} \; ; \; (\boldsymbol{H}_{sca} + \boldsymbol{H}_{inc} - \boldsymbol{H}_c) \times \boldsymbol{e}_r = \boldsymbol{0}$$

(6)

We can therefore set up an inhomogeneous system of linear equations $A\boldsymbol{x} = \boldsymbol{u}$ where $A$ is an $8 \times 8$

$$\boldsymbol{u} = \begin{bmatrix} 0 \\ 0 \\ 0 \\ 0 \\ -\dfrac{k_M^2 - h_M^2}{k_M^2} J_n[\rho_M(r_s)] \\ -\dfrac{k_M^2 - h_M^2}{k_M^2} \dfrac{J_n[\rho_M(r_s)]}{\rho_M(r_s)} - i\dfrac{\sqrt{k_M^2 - h^2}}{k_M} J_n'[\rho_M(r_s)] \\ \dfrac{(k_M^2 - h_M^2)}{k_M} J_n[\rho_M(r_s)] \\ \dfrac{\sqrt{k_M^2 - h_M^2}}{k_M} hn \dfrac{J_n[\rho_M(r_s)]}{\rho_M(r_s)} + i\sqrt{k_M^2 - h_M^2} J_n'[\rho_M(r_s)] \end{bmatrix} \; ; \; \boldsymbol{x} == \begin{bmatrix} a_n \\ b_n \\ \alpha_n^c \\ \beta_n^c \\ \alpha_n^s \\ \beta_n^s \\ \gamma_n^s \\ \delta_n^s \end{bmatrix}$$

(7)

matrix for the unpolarised case, $\boldsymbol{x}$ is a $8 \times 1$ vector of the expansion coefficients defined earlier and $\boldsymbol{u}$ is a constant vector. $\boldsymbol{x}$ and $\boldsymbol{u}$ are given in equation (7).

The matrix A is given in full detail in the equation below while considering a random incident wave of angle $\xi$.

$$A = \begin{bmatrix} 0 & 0 & -A_{13}^{c,c,J_n} & 0 & A_{15}^{s,c,J_n} & A_{16}^{s,c,Y_n} & 0 & 0 \\ 0 & 0 & -B_{23}^{c,c,J_n} & -C_{24}^{c,c,J_n} & B_{25}^{s,c,J_n} & iC_{26}^{s,c,J_n} & B_{27}^{s,c,Y_n} & iC_{28}^{s,c,Y_n} \\ 0 & 0 & 0 & D_{34}^{c,c,J_n} & 0 & -D_{36}^{s,c,J_n} & 0 & -D_{38}^{s,c,Y_n} \\ 0 & 0 & ik_c C_{43}^{c,c,J_n} & k_c B_{44}^{c,c,J_n} & -ik_s C_{45}^{s,c,J_n} & -k_s B_{46}^{s,c,J_n} & -ik_s C_{47}^{s,c,J_n} & -k_s B_{48}^{s,c,Y_n} \\ A_{51}^{M,s,H_n} & 0 & 0 & 0 & -A_{55}^{s,c,J_n} & 0 & 0 & A_{58}^{s,s,Y_n} \\ B_{61}^{M,s,H_n} & ik_M C_{62}^{M,s,H_n} & 0 & 0 & -B_{65}^{s,s,J_n} & -ik_s C_{66}^{s,s,J_n} & -B_{67}^{s,s,Y_n} & -ik_s C_{68}^{s,s,Y_n} \\ 0 & -k_M A_{72}^{M,s,H_n} & 0 & 0 & 0 & k_s A_{76}^{s,s,J_n} & 0 & k_s A_{78}^{s,s,Y_n} \\ -ik_M C_{81}^{M,s,H_n} & -k_M B_{82}^{M,s,H_n} & 0 & 0 & k_s C_{85}^{s,s,J_n} & k_s B_{86}^{s,s,J_n} & ik_s C_{87}^{s,s,Y_n} & k_s B_{88}^{s,s,Y_n} \end{bmatrix}$$

where each entry of the matrix A is described in detail in equation (8).

$$A_{pq}^{i,j,Z_n} = \frac{k_i^2 - h_i^2}{k_i^2} Z_n[\rho_i(r_j)]; \ B_{pq}^{i,j,Z_n} = \frac{\sqrt{k_i^2 - h_i^2}}{k_i^2} hn \frac{Z_n[\rho_i(r_j)]}{\rho_i(r_j)}$$

$$C_{pq}^{i,j,Z_n} = \frac{\sqrt{k_i^2 - h_i^2}}{k_i} Z_n'[\rho_i(r_j)]; \ D_{pq}^{i,j,Z_n} = \frac{(k_i^2 - h_i^2)}{k_i} Z_n[\rho_i(r_j)]$$

$$i,j \in \{(M,c,s) \times (c,s)\}; Z_n \in \{J_n, Y_n, H_n\}$$

(8)

It is important to mention that the matrix A is ill-conditioned, which implies the non-uniqueness of solutions for the equation $A.x = u$. Nevertheless, it is possible to solve it using GMRES algorithm already available in MATLAB® after improving the ill-conditioned matrix A using the method described in Ref. [10] .It is worth noting that the novelty of this approach resides in the explicit dependence of the TM and TE modes with the incidence angle $\xi$.

Furthermore, after we have obtained the electromagnetic fields inside the core and the shell as well as the scattered fields by the cylindrical core-shell nanowire, we can determine the Poynting vector at any point in space as a possible way to quantify energy distribution inside the core-shell structure, we evaluated the normalized time-averaged Poynting vector defined as

$$\langle S \rangle = Re[E \times H^*] \tag{9}$$

We focused on the TE-TM polarization of the Poynting vector but we presented only the TM-like mode, due to low scattering and absorption efficiencies of TE-like mode at small core radii, where the highest efficiencies are usually located.

Finally, the analysis for the $Si/Ge_{0.88}Sn_{0.12}/Si_3N_4$ NW structure is similar to the CSNW one. Nevertheless, the differences will emerge when writing the EM field continuity equation around the three different interfaces: core/ inner-shell, inner-shell/outer-shell and outer-shell/medium (air). The mathematics development will not be shown here because it is similar to the CSNW case.

## S3.    Experimental Validation of the Mie-Lorentz Calculation scattering for SiNW

It is detrimental to evaluate the correctness and the exactitude of the Mie-Lorentz scattering approach before endeavouring in more complicated computation. For that reason, we start by studying the simplest system of SiNW by calculating the absorption or scattering efficiencies and comparing the results to the experimental measurement or theoretical calculation, reported in literature whenever available. Initially, we validate our calculation by reproducing the TE-polarized scattering efficiency of the SiNW based on the work of R. P-Dominguez *et al.*[11] The results are shown in Figure S2A where we can see that we accurately reproduce the same resonance peaks for a silicon cylinder with a 170 nm radius as in Figure 1 from the previous reference. Additionally, we replicate the experimental measurement of scattering efficiencies of silicon nanowires with different diameters ranging from 30 to 180 nm, presented in the Figure 2b in the work of Cao *et al.*[12]. However, we need to verify that the approximation of infinitely long nanowire used in the Mie-Lorentz formalism still hold, where the diameter to the length ratio need to be negligible before $10(d/L \ll 10)$. For instance, by inspecting the Figure 1b from the aforementioned reference, we can deduce that the length of the NW is approximately 8 µm and by comparing it to the biggest diameter (180 nm) from Figure 2b [12], we can deduce that $d/L = 0.0225 \ll 10$. Thus, we can ascertain that the approximation of infinitely long nanowire still holds. Next, we present the result of our simulation in Figure S2B

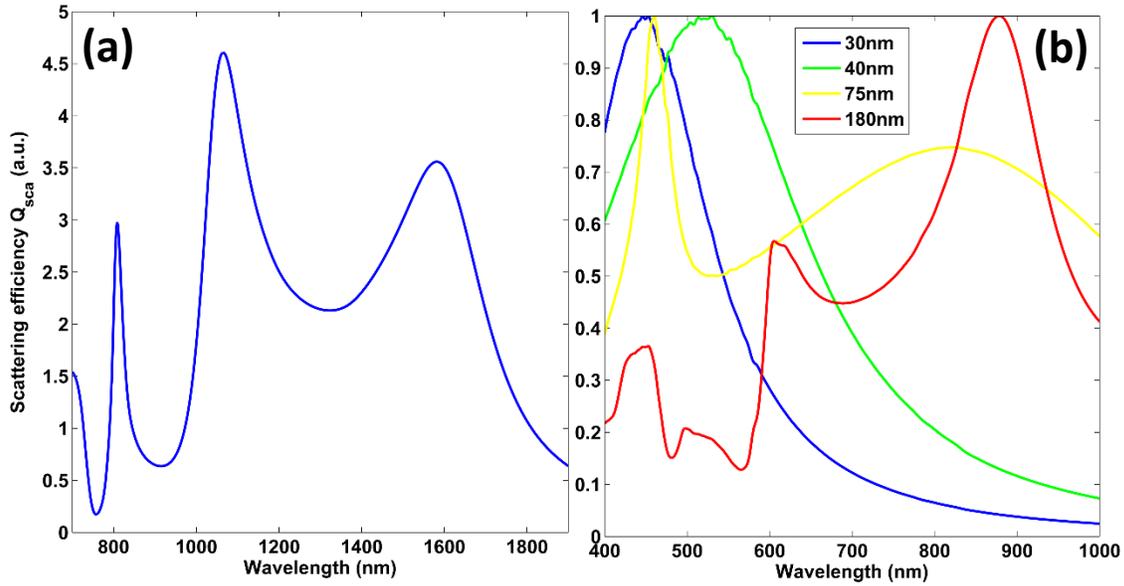

**Supplementary Figure S2:** Scattering efficiency of silicon nanowire **(a)** with a radius of 170 nm under TE-polarized plane wave incidence based on Ref. [11] and **(b)** with different diameters of 30 nm (blue), 40 nm (green), 75 nm (yellow), and 180 nm (red) under randomly polarized light based on Ref. [12] calculated with the Mie-Lorentz scattering formalism. The $Q_{sca}$ was normalized to highlight spectral changes.

## S4.    Full-Vector Finite-Difference complex mode solver for CSNW

In order to better understand the correlation between the geometrical dimension of the CSNW (the core radius and the shell thickness) and the absorption, we present in Fig. S3a, we fix the core radius to 60 nm and we vary the shell thickness from $0.25 \times R_c$ to $R_c$. We can clearly see the Leaky mode $TM_{41}$ highlighted with an eclipse. When the shell thickness increases, the leaky resonance mode moves toward longer wavelength, thus the polarized light is redshifted.

In addition, to better analyze the effect of the geometrical dimensions on the leaky modes of the CSNW, we used the improved full-vector Finite Difference (FVFD) mode solver for general circular waveguides, proposed in References [13,14], coupled with the coordinate stretching technique to implement the Perfectly Matched Layer (PML) boundary condition, in order to accurately extract and distinguish the leaky modes from the guided ones. Thus, we find the complex effective refractive index $n_{eff}$ of the CSNW. Each complex solution is the eigenvalue of a specific leaky mode. The real part ($N_{real}$) of the eigenvalue is related with the wavelength where the optical resonance (LMRs) takes place ($\lambda = 2\pi r_c/N_{real}$), and the imaginary part ($N_{imag}$) dictates the spectral width of the optical resonance. The Leaky waveguide modes are classified based on their azimuthal mode number, $m$, their radial mode number, $n$, which arises from the oscillatory behavior of the Bessel functions, and their polarization. Their polarization can be either TM (transverse magnetic, $H_z=0$), TE (transverse electric, $E_z = 0$), HE (magnetoelectric, TM-like), or EH (electromagnetic, TE-like). The only modes that are strictly TE or TM for arbitrary wavevector are the $0^{th}$ order azimuthal modes, $TM_{0n}$ and $TE_{0n}$. From the field profiles, presented in Supplementary Figure S3b, it is clear that the peaks at the 354 nm and 644 nm can be attributed to the $HE_{11}$ and $TM_{01}$ leaky waveguide modes, respectively when $R_c = 8\ nm$ and $t = 0.25R_c$. When $R_c = 60\ nm$, there are 3 main modes that are present: $TM_{21}$, $TM_{31}$ and $TM_{41}$ having respectively the following resonant wavelength: 988.3 nm, 851.7 nm and 681.2 nm. The field profiles of these modes, as determined from FVFD, can be found in references [15–17]. From absorption and scattering spectra, it can be observed that the lower modes broaden in scattering spectra and higher orders tend to disappear entirely, especially for smaller sizes (red curve of Figure S3b), whereas for larger size, higher orders appear at near infrared region of the spectrum.

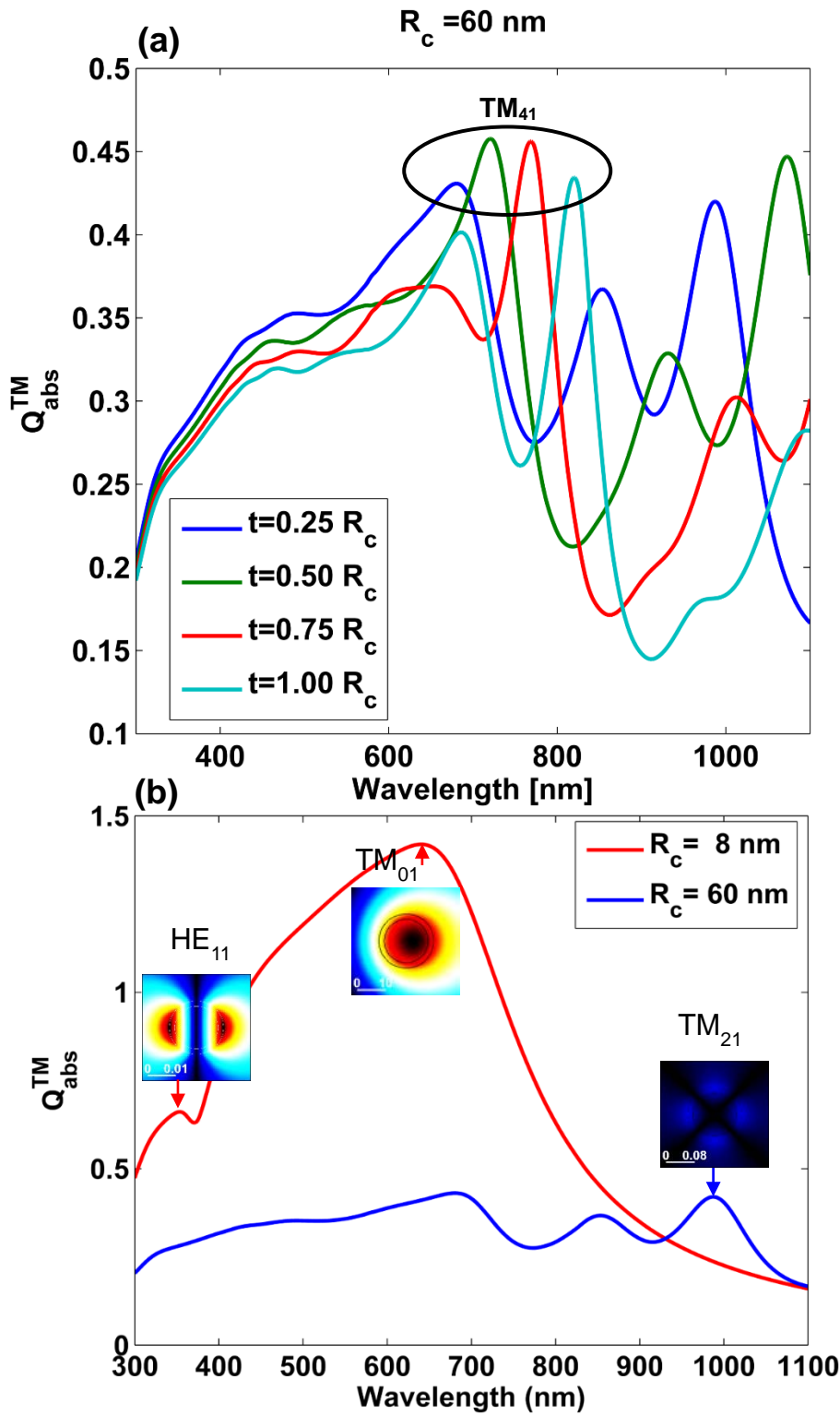

**Supplementary Figure S3: (a)** Effect of the shell thickness $t$ on the leaky mode spectral position for the Si/GeSn CSNW. We specifically highlight the correlation between the shell thickness and the $TM_{41}$ leaky mode for a core radius of 60 nm. Next, we present in panel **(b)** TM polarized absorption efficiency of Si/GeSn CSNW for 2 distinct core radii of 8 nm (red curve) and 60 nm (blue curve) and a shell thickness of $t = 0.25R_c$ as well as some corresponding electric field profile at the corresponding wavelength 354 (HE$_{11}$) and 644 nm (TM$_{01}$) for $R_c$=8 nm and 989 nm (TM$_{21}$) for $R_c$=60 nm.

## S5.    Light Absorption Enhancement in Ge/Ge$_{0.88}$Sn$_{0.12}$ and Si/Ge$_{0.84}$Si$_{0.04}$Sn$_{0.12}$ CSNWs

We evaluate light absorption enhancement $\eta_{abs}$ relative to SiNW for Si/Ge$_{0.88}$Si$_{0.04}$Sn$_{0.12}$ CSNW and relative to GeNW for Ge/Ge$_{0.88}$Sn$_{0.12}$ CSNW. We show below in Supplementary Figure S3 the extreme enhancement for Si based CSNW in the NIR region. It is clear that the Ge/Ge$_{0.88}$Sn$_{0.12}$ CSNW is a less efficient light absorber than the Si/Ge$_{0.88}$Si$_{0.12}$Sn$_{0.04}$ CSNW. We get for Ge/Ge$_{0.88}$Sn$_{0.12}$ CSNW at specific core radius 31 nm and shell thickness ($t = R_c$) and at a wavelength of 1100 nm, a 12-fold absorption enhancement compared to the GeNW.

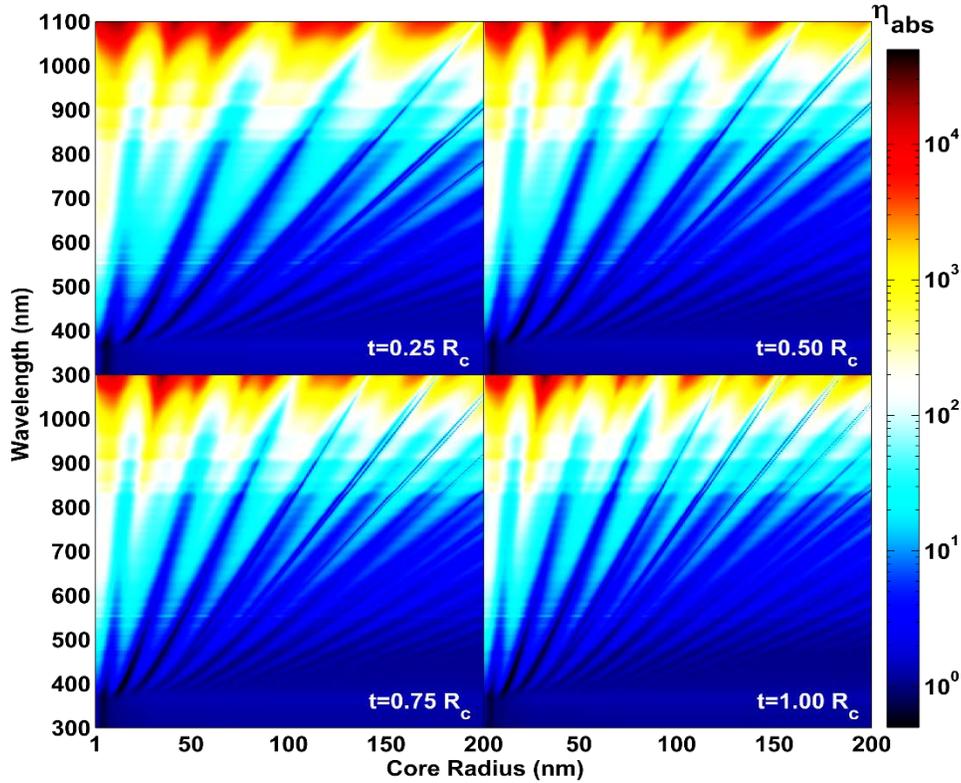

**Supplementary Figure S4:** Extreme light absorption enhancement map as a function of the core radius and the incident light wavelength for the Si/Ge$_{0.84}$Si$_{0.04}$Sn$_{0.12}$ CSNW for different shell thicknesses $t$= [0.25, 0.5, 0.75, 1] ×R$_c$.

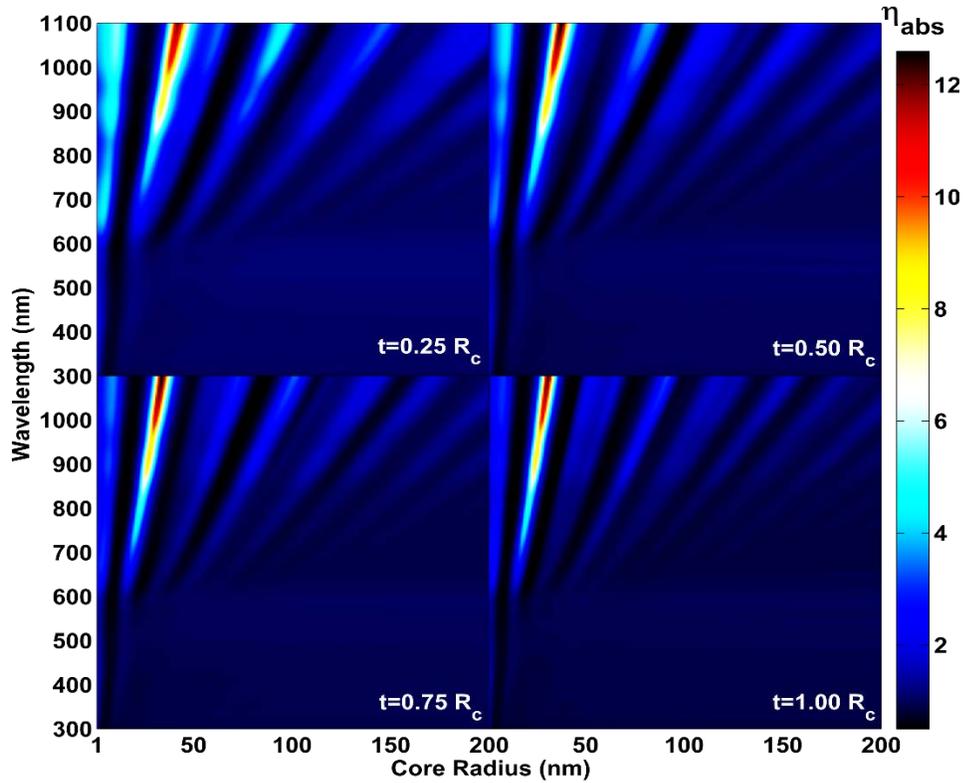

**Supplementary Figure S5:** Light absorption enhancement map as a function of the core radius and the incident light wavelength for different core-dependent shell thicknesses for the Ge/Ge$_{0.88}$Sn$_{0.12}$ core-shell nanowire.

# Supplementary File References: